\documentclass[twocolumn]{aastex61}
\usepackage{amsmath}
\usepackage{longtable}
\usepackage{multirow}

\newcommand{\mathsc}[1]{{\text{\normalfont\scshape#1}}} 
\newcommand\ho{\ifmmode {\rm H\hspace{.1em}\mathsc{i}} \else \mbox{\rm H\,\scshape{i}} \fi}
\newcommand{\kms}{\ifmmode {\rm km\ s}^{-1} \else km s$^{-1}$ \fi}
\newcommand{\oii}{[O {\sc ii}]}
\newcommand{\oiii}{[O {\sc iii}]}
\newcommand{\nii}{[N {\sc ii}]}
\newcommand{\msun}{M_{\odot}}
\newcommand{\hi}{\mbox{\rm H$\,$\scshape{i}}}

\begin{document}

\shorttitle{The physical properties of S0 galaxy}
\title{The physical properties of S0 galaxy PGC 26218: the origin of starburst and star formation }

\correspondingauthor{Qiu-Sheng Gu}
\email{Email: qsgu@nju.edu.cn}

\author{Xue Ge} 
\affil{School of Astronomy and Space Science, Nanjing University, Nanjing, Jiangsu 210093, China}
\affiliation{Key Laboratory of Modern Astronomy and Astrophysics (Nanjing University), Ministry of Education, Nanjing 210093, China}
\affiliation{Collaborative Innovation Center of Modern Astronomy and Space Exploration, Nanjing 210093, China}

\author{Qiu-Sheng Gu$^{\color{blue} \dagger}$} 
\affil{School of Astronomy and Space Science, Nanjing University, Nanjing, Jiangsu 210093, China}
\affiliation{Key Laboratory of Modern Astronomy and Astrophysics (Nanjing University), Ministry of Education, Nanjing 210093, China}
\affiliation{Collaborative Innovation Center of Modern Astronomy and Space Exploration, Nanjing 210093, China}

\author{Rub\'en Garc\'ia-Benito} 
\affil{Instituto de Astrof\'isica de Andaluc\'ia (CSIC), P.O. Box 3004, 18080 Granada, Spain}

\author{Meng-Yuan Xiao} 
\affil{School of Astronomy and Space Science, Nanjing University, Nanjing, Jiangsu 210093, China}
\affiliation{Key Laboratory of Modern Astronomy and Astrophysics (Nanjing University), Ministry of Education, Nanjing 210093, China}
\affiliation{Collaborative Innovation Center of Modern Astronomy and Space Exploration, Nanjing 210093, China}

\author{Zong-Nan Li} 
\affil{School of Astronomy and Space Science, Nanjing University, Nanjing, Jiangsu 210093, China}
\affiliation{Key Laboratory of Modern Astronomy and Astrophysics (Nanjing University), Ministry of Education, Nanjing 210093, China}
\affiliation{Collaborative Innovation Center of Modern Astronomy and Space Exploration, Nanjing 210093, China}

millimeter

\begin{abstract}

We present 2D-spectroscopic observations from Centro Astron\'omico Hispano Alem\'an (CAHA) 3.5 m telescope and the millimetre observation from NOrthern Extended Millimeter Array (NOEMA) of the nearby S0 galaxy PGC 26218, which shows central star-formation activity and post-starburst outside in the disk. We estimate the star formation rate (SFR = $0.28\pm0.01$ $M_{\odot} \rm yr^{-1}$) and molecular gas mass 
($\rm log\ $$M_{\rm H_{2}}=7.60\pm0.15\ M_{\odot}$) of PGC 26218 based on the extinction-corrected H$\alpha$ emission line and the CO-$\rm H_{2}$ conversion factor ($\alpha_{\rm CO}$) of the Milky Way, respectively. We find that PGC 26218 follows the star forming main sequence (SFMS) and the Kennicutt-Schmidt law. Comparing the kinematics of CO($J$=1-0), stars and H$\alpha$, we find that the rotational axis of CO($J$=1-0) is 45$^{\circ}$ different from that of H$\alpha$. In addition, the profile of the CO($J$=1-0) emission line shows asymmetry and has an inflow component of $\sim$ 46 $\rm km\ s^{-1}$. With the decomposition of the optical image, we confirm that PGC 26218 shows multiple nuclear structures. The projected offset between the most luminous optical center and the center of CO($J$=1-0) is $5.2\arcsec$ ($\sim$ 0.6 kpc) and the latter overlaps with one of the optical cores.  These results support that PGC 26218 may have experienced a gas-rich minor merger, extending its star formation and locating it in the SFMS.

\end{abstract}

\keywords{galaxies: starburst --- galaxies: star formation --- galaxies: elliptical and lenticular, cD --- galaxies: peculiar}

\section{Introduction}
\label{sec1}

Lenticular galaxies (S0s) are typically considered to be the intermediate transition population between spiral and elliptical galaxies in the Hubble tuning fork diagram \citep{1936Sci....84..509M}. They are classified as early-type galaxies (ETGs) based on the morphology. The prominent features in S0 galaxies are the absence of spiral arms and noticeable star formation regions.

Some forming scenarios on the S0 galaxies have been proposed.  On the one hand, S0 galaxies can be formed by the morphological transformation of spiral galaxies, which consume the gas in their disks. \citep{2007astro.ph..2125B, 2011MNRAS.415.1783B, 2012ApJS..198....2K, 2014ASPC..480..161J, 2018MNRAS.476.2137R}. 
On the other hand, the external effects, such as galaxy harassments \citep{1996Natur.379..613M}, tidal encounters in high-density environments and cluster gravitational potential \citep{1998ApJ...502L.133B, 1998ApJ...495..139M, 2006A&A...458..101A, 2009MNRAS.398..312G, 2010MNRAS.405.1089L}, and galaxy mergers  \citep{2001A&A...367..428A, 2006A&A...457...91E, 2015A&A...573A..78Q, 2017A&A...604A.105T} may also lead to the formation of S0 galaxies. 
In addition, The existence of a bar may also play an important role in the formation of S0 galaxies via gas transport to the center, fuelling a starburst \citep{2006AJ....132.2634L}.
A recent study has shown that violent disk instability could also be an important forming mechanism of S0 galaxies \citep{2018ApJ...862L..12S}. 
Generally, S0 galaxies are gas-poor, inactive galaxies. They likely formed through a combination of multiple processes due to the diverse properties in their bulges and disks \citep{2010MNRAS.405.1089L, 2013MNRAS.432..430B, 2018MNRAS.481.5580F} and help us understand the diverse evolutionary paths from star-forming blue galaxies to quiescent red galaxies.

Although S0 galaxies are often thought to have evolved passively since a big burst of star formation, some studies found that these galaxies still have nuclear star-formation activity \citep{2007ApJS..173..619K, 2007MNRAS.382.1415S}. Further studies have found that the neutral and even molecular gas is present in most of S0 galaxies \citep{1991A&A...243...71V, 2003ApJ...584..260W, 2006ApJ...644..850S, 2010ApJ...725..100W}. This reservoir of gas could be explained via the gas-rich mergers \citep{2015MNRAS.449.3503D} and gas accretion from the environment \citep{2013ApJ...770...62D}. In addition, N-body simulations show that the stellar mass loss can also fuel the residual star formation in massive ETGs \citep{2001A&A...376...85J}.

\begin{figure}
\centering
\includegraphics[angle=0,width=0.35\textwidth]{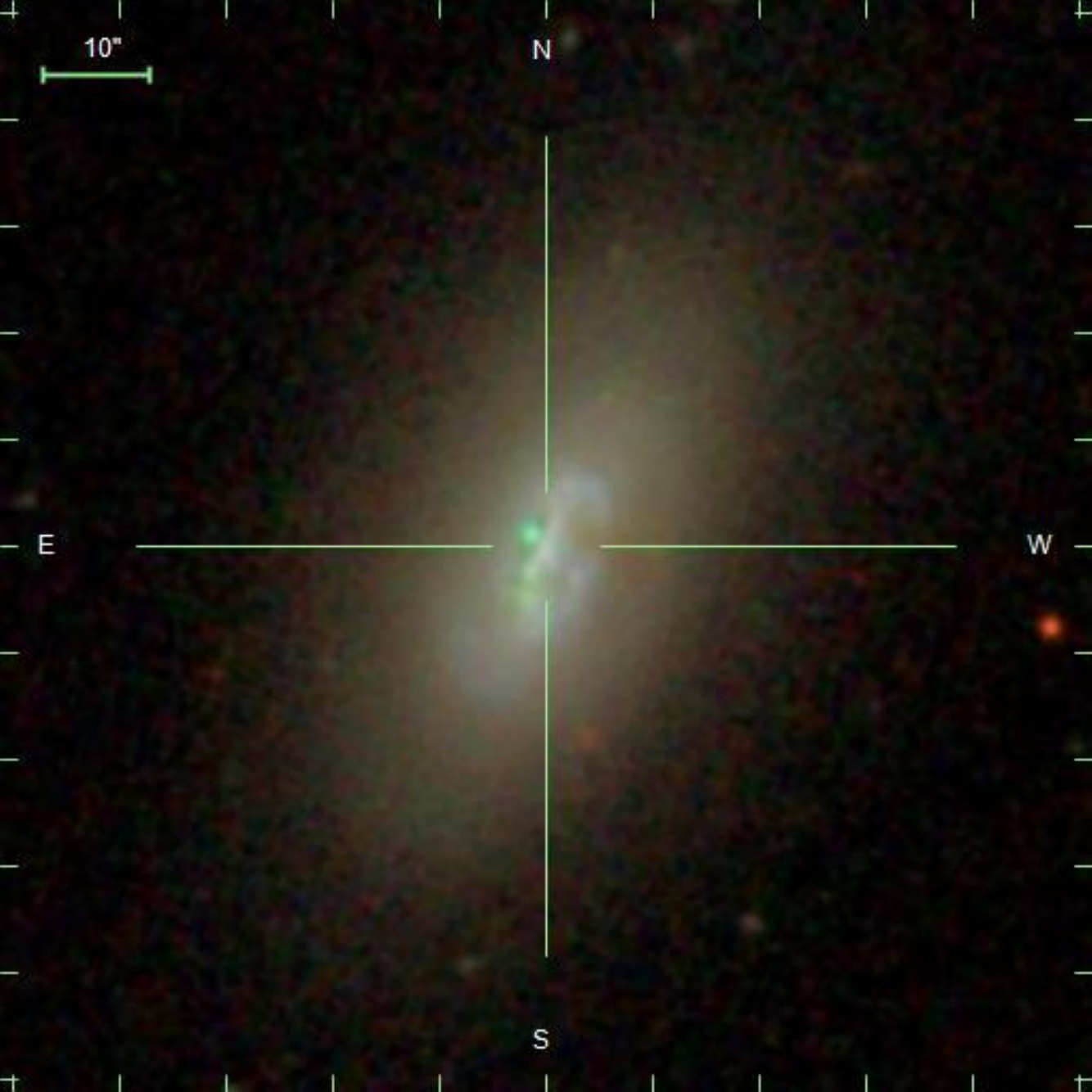}
\caption{SDSS $gri$ image of PGC 26218 and $\rm 1\arcsec \sim 0.11\ kpc$.}
\label{f1}
\end{figure}

Based on the revised Third Reference Catalog of Bright Galaxies \citep[RC3,][]{2015arXiv151201204L} and Sloan Digital Sky Survey (SDSS) Data Release 7 \citep{2009ApJS..182..543A},  \cite{2016ApJ...831...63X} collected a visual morphology sample of S0 galaxies. They presented the properties of nuclear activities for the sample of S0 galaxies and found that 45 (8 percent) nearby S0 galaxies show signs of nuclear star-formation activity.
In order to understand the nature and spatially resolved properties of this sample of star-forming S0 galaxies, we have started a program to obtain Integral Field Spectroscopy data with the Centro Astron\'omico Hispano Alem\'an (CAHA) 3.5 m telescope. SDSS J091705.28+252545.4 (PGC 26218) is one of the observed star-forming S0 galaxies ($z$ = 0.00548). Figure \ref{f1} shows the SDSS composite color image. We can see the disturbed structure and a bright star-forming knot in the center of PGC 26218.
In order to investigate the molecular gas in this S0 galaxy, we observed the CO($J$=1-0) emission line with NOrthern Extended Millimeter Array (NOEMA). Our main purposes in this paper is to determine the origins of star formation in PGC 26218 and whether such a S0 galaxy follows the star formation laws as normal star-forming galaxies. 

The paper is organized as follows. Section \ref{sec2} and \ref{sec3} show the observations and data analysis of optical  Integral Field Unit (IFU) and millimeter. Section \ref{sec4} gives our results and discussions, including the star formation in PGC 26218, star forming main sequence (SFMS), Kennicutt-Schmidt (K-S) law, the kinematics of star and gas, and the origins of star formation. In Sections \ref{sec5}, we present our summary. Throughout the paper, we adopt a cosmology with $\Omega_{M}$=0.3, $\Omega_{\Lambda}$=0.7 and $\rm H=70~km~s$$^{-1}$~Mpc$^{-1}$ and \cite{1955ApJ...121..161S} IMF.

%
\begin{center}
\begin{table}[h]
\begin{footnotesize}
\renewcommand{\thetable}{\arabic{table}}
\caption{The archival parameters of PGC 26218. }
\label{tab1}
\begin{tabular}{lr}
\tablewidth{0pt}
\hline
\hline
PGC 26218		&  						\\
\hline
R.A. (SDSS) [J2000.0] 			& 	139.272017					\\
Decl. (SDSS) [J2000.0]			& 	25.429165 				\\
Redshift (SDSS)		                &	0.00548						\\
$\log$ $M_*$ [$M_{\sun}$] (MPA-JHU DR7\footnote{https://wwwmpa.mpa-garching.mpg.de/SDSS/DR7/})	   &	$9.32^{\rm s}/9.15^{\rm k}$\\
\hline
\end{tabular}
\end{footnotesize}
\tablecomments{The superscript ``s'' on the stellar mass indicates that the stellar mass was computed adopting \cite{1955ApJ...121..161S} IMF, while ``k''  \cite{2001MNRAS.322..231K} IMF.}
\end{table}
\end{center}

%
%
\section{Observations and Data Reductions}
\label{sec2}

In this section, we present our data observations and reductions of optical spectroscopy and millimeter, respectively. Based on the preliminary analysis, we aim to give the reader a general perception of the nearby S0 galaxy PGC 26218. The archival parameters and the results from the data analysis for PGC 26218 are summarized in Table \ref{tab1} and Table \ref{tab2}, respectively.

\subsection{CAHA 2D-Spectroscopic Observation}
\label{sec2.1}

We have obtained optical IFU spectroscopic observations of PGC 26218 from CAHA on 2016 March 11 and 12. The 3.5-m telescope covers two optical overlapping setups. The red spectrograph covers the wavelength range from 3745 to 7500 \AA\ with a low spectral resolution (R $\sim$ 850, V500), while the blue covers the wavelength range from 3400 to 4840 \AA\ with a medium spectra resolution (R $\sim$ 1650, V1200). In order to obtain a filling factor of 100\%, a 3-pointing dithering scheme was used. The exposure time per pointing was of 900s for the blue (taken on the first day) and 1800s for the V1200 (split in 2 individual exposures of 900s). We used a python-based pipeline for reduction of the PPAK data based on an upgraded version of \cite{2015A&A...576A.135G} and \cite{2016A&A...594A..36S}. The reduction process can be summarized in the following steps: identification of the position of the spectra on the detector along the dispersion axis; extraction of each individual spectrum; distortion correction of the extracted spectra; wavelength calibration; fiber-to-fiber transmission correction; flux-calibration; sky-subtraction; cube reconstruction; and finally differential atmospheric correction. To reduce the effects of the vignetting on blue part (up to $\sim$ 4200 \AA) of the V500 data in some parts of the FoV (see Fig. 11 of \cite{2013A&A...549A..87H}), we combine both setups producing a so-called COMBO datacube. The V1200 spectral resolution is degraded to match the V500 data. We then combine the data from both datasets averaging the V1200 cube and V500 cube in the overlapping wavelength range, weighted by the inverse of the error. For the remaining wavelength range, the blue part corresponds to the matched V1200 cube and the red part, to the right of the overlapping region, to the original V500 datacube. More details of the reduction process can be found in \cite{2013A&A...549A..87H}, \cite{2015A&A...576A.135G} and \cite{2016A&A...594A..36S}. The final datacube, corrected for Galactic extinction, fully covers the optical range from 3700 to 7300 \AA. 

\subsection{The Fitting of Stellar Continuum and Emission Lines}
\label{sec2.2}

Preprocessed spectra data (78 $\times$ 73 spaxels) are stored in a 3D data cube. We firstly bin the 2D spectra data to a constrained signal-to-noise ratio (SNR) with the Voronoi binning method \citep{2003MNRAS.342..345C}. We adopt the average flux and the standard deviation of the flux as the signal and noise at the wavelength range from 5075 \AA\ to 5125 \AA. This range is not contaminated by the emission and absorption lines. Spaxels below a lower SNR limit set by us are stacked together neighbour spaxels until they satisfy the required minimum SNR. We tested several realizations with SNR = 5, 10, 20 and finally adopted SNR = 10 in this work. Although higher SNR is better for the fitting, this will reduce the number of spaxels and smooth the details in space. This process produces 1370 Voronoi bins with SNR higher than 10, 80\% of which include only one spaxel. Next, we use pPXF code \citep{2004PASP..116..138C, 2017MNRAS.466..798C}, a stellar population analysis method, to fit the stellar continuum, during which the emission lines are masked (such as H$\alpha$, \oiii, H$\beta$, etc). The code adopts MILES simple stellar population templets \citep{2006MNRAS.371..703S, 2010MNRAS.404.1639V}, assuming a \cite{1955ApJ...121..161S} IMF and \cite{2000ApJ...533..682C} dust extinction curve, covering 25 population ages (between 0.06 and 15.85 Gyr) and 6 metallicities (log[M/H]=-1.71, -1.31, -0.71, -0.4, 0.0, 0.22). All spectra are fitted in the wavelength range from 3800 to 7200 \AA. We check visually all the fitting results and excluded some peculiar fits with stellar velocity dispersion higher than 400 $\rm km\ s^{-1}$ and $D_{n}4000$ (the ratio of continuum 4000-4100 \AA\ and 3850-3950 \AA, \cite{1999ApJ...527...54B}) higher than 2.0, which results in 1200 spectra analyzed in this work. We emphasize that the elimination of spaxels by the visual inspection process does not significantly affect our results.

Figure \ref{f2} shows the distributions of fitted parameters derived from pPXF. Panel A represents the $D_{n}4000$, an indicator for stellar population age \citep{2003MNRAS.341...33K}. The small value of $D_{n}4000$ implies that there is on-going star formation in the central region. The similar trend can be found from the light-weighted age (panel B), which is weighted by the fraction of light for each template. Panel C represents the stellar velocity map, showing a characteristic rotated disk-shape structure. It is interesting that the equivalent width (EW, positive value represents absorption) of H$\delta$ line is large along the major axis (panel D), suggesting that these regions might experienced a starburst process several hundred Myr ago. 

For the emission lines, we subtract the estimated stellar continuum from the observed spectrum and obtain the line flux with the SHERPA IFU line fitting software (SHIFU; Garc\'ia- Benito, in preparation), based on the package of CIAO SHERPA \citep{2001SPIE.4477...76F, 2007ASPC..376..543D}. Small deviations with respect to the stellar continuum are taken into account by a first-order polynomial. Single Gaussians have been fitted for the emission lines and the width of the Gaussian were tied for ions of the same element.

\subsection{The Selection of Regions with Different Star Formation Histories}
\label{sec2.3}

PGC 26218 is a star-forming S0 galaxy and displays the post-starburst outside in the disk. In order to show the locations of regions with different star formation histories, we pick out these regions according to the EW of H$\alpha$ and H$\delta$.

A traditional picture about post-starburst galaxies is that these objects have experienced violent starburst and then rapidly quenched star formation within the last few hundred Myr. Their optical spectra are characterized by strong Balmer absorption lines and weak nebular emission lines. Therefore, the post-starburst galaxies are usually selected based on the deficiency of emission lines (such as H$\alpha$ or \oii) and the strong H$\delta$ absorption line  \citep{2004MNRAS.355..713B, 2004ApJ...602..190Q, 2005MNRAS.357..937G, 2007MNRAS.381..187G, 2009ApJ...693..112P, 2010A&A...509A..42V}. The strong high-order Balmer absorption lines evidence the existence of A-class stars, while the absence of emission lines indicates that there is no recent star formation in the past Gyr.

\begin{figure*}
\centering
\includegraphics[angle=0,width=0.8\textwidth]{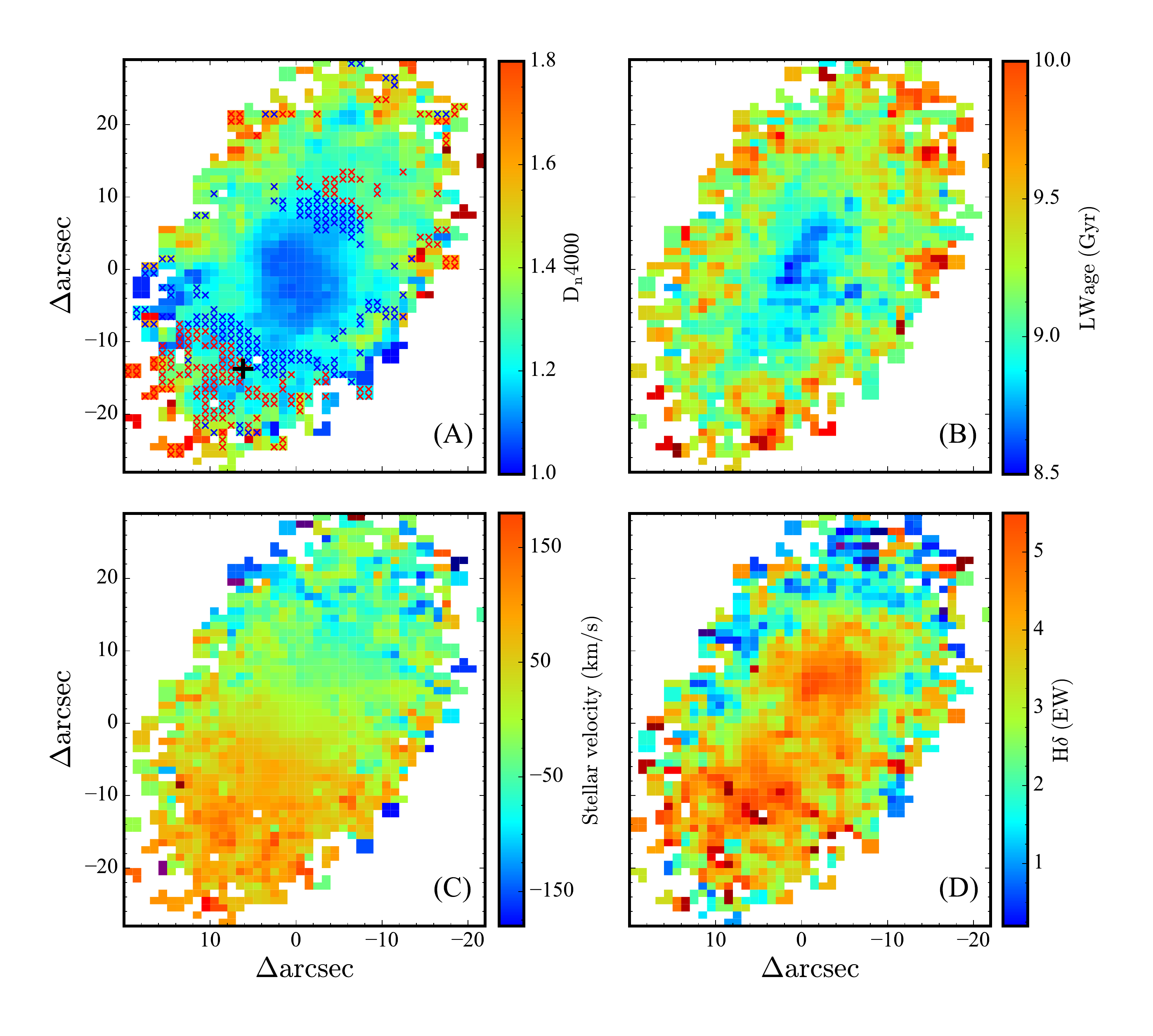}
\caption{The distributions of optical parameters color coded by $D_{n}4000$ (A), light weight age (B), velocity (C), and EW (H$\delta$, D), respectively. The positive value in EW denotes the absorption. The map center is set at the RA, DEC = 139.272, 25.429 and 1 arcsec $\sim$ 0.11kpc. In panel A, the red and blue crosses represent quiescent post-starburst and transiting post-starburst regions, respectively based on the EW of H$\alpha$ and H$\delta$.
The black cross indicates the center of millimeter observation.}
\label{f2}
\end{figure*}

The traditional definition for post-starburst imposes a rigorous cut on H$\alpha$ emission line \citep{2005MNRAS.360..587B, 2006ApJ...650..763H}. The restriction on H$\alpha$ intensity may miss some post-starburst galaxies that reside in the early evolutionary phase. Recent studies show that the early stage is also important for understanding the evolution of post-starburst galaxies \citep{2014ApJ...792...84Y, 2015MNRAS.448..258R, 2016ApJS..224...38A, 2018MNRAS.477.1708P}. In this work, the quiescent post-starburst regions (QPSB) are defined as $\rm EW (H\delta) > 4\ \AA$ and $\rm EW (H\alpha) > -3\ \AA$, while the transitioning post-starburst regions (TPSB) are defined as $\rm EW (H\delta) > 4\ \AA$ and $\rm -10\ \AA < EW (H\alpha) < -3\ \AA$. For QPSB, the definition is the same as \citet{2014ApJ...792...84Y}. The upper limit on $\rm EW (H\alpha)$ ensures that the TPSB region has only residual star formation. The typical errors for the EW of H$\alpha$ and H$\delta$ are 1.0 \AA\ and 1.4 \AA, respectively.
The corresponding regions are marked in panel A of Figure \ref{f2}. We find that the QPSB regions (red crosses) are mainly located outside 1.5 effective radius ($R_{e} \sim 13\arcsec$), while the TPSB regions (blue crosses) mainly concentrated around the center of PGC 26218. Furthermore, we define the regions with $\rm EW (H\delta) < 4\ \AA$ and $\rm EW (H\alpha) < -3\ \AA$ or $\rm EW (H\delta) > 4\ \AA$ and $\rm EW (H\alpha) < -10\ \AA$ as the star-forming (SF) regions. We emphasize that the number of spaxels for QPSB accounts for 20\% of the total spaxels of these three regions. This significative amount of QPSB spaxels shows that PGC 26218 has undergone a starburst several hundred Myr ago. Based on the optical features, millimeter observation coordinates (black cross) is selected between the two PSB regions. We will describe the CO($J$=1-0) emission in Section \ref{sec3}.

\begin{figure*}
\centering
\includegraphics[angle=0,width=0.8\textwidth]{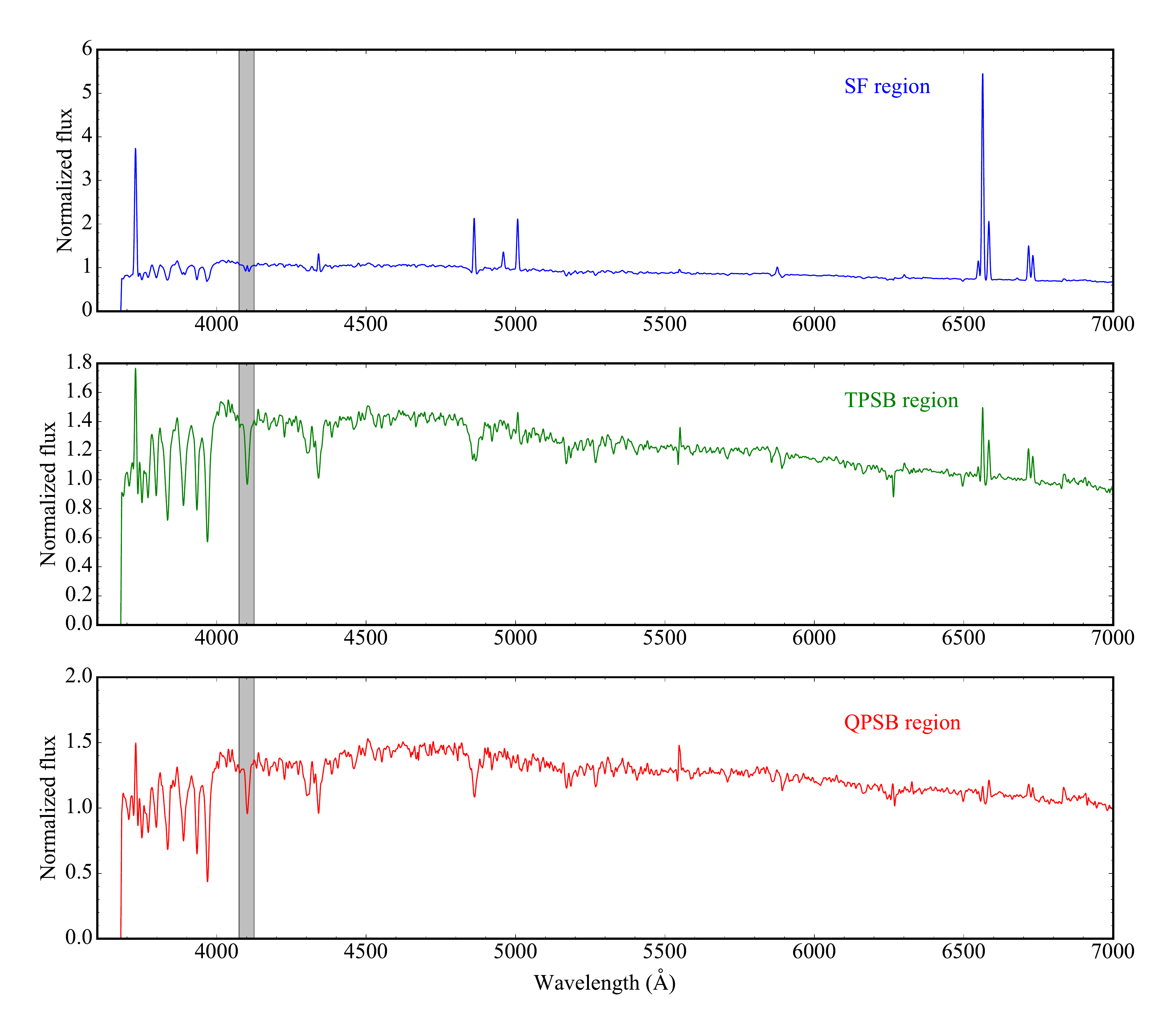}
\caption{Stacked spectra for SF (top), TPSB (middle) and QPSB (bottom), respectively. All the spectra are normalized at 4100 \AA. The shadow marks the region of H$\delta$.}
\label{f3}
\end{figure*}

\begin{figure*}
\centering
\includegraphics[angle=0,width=0.8\textwidth]{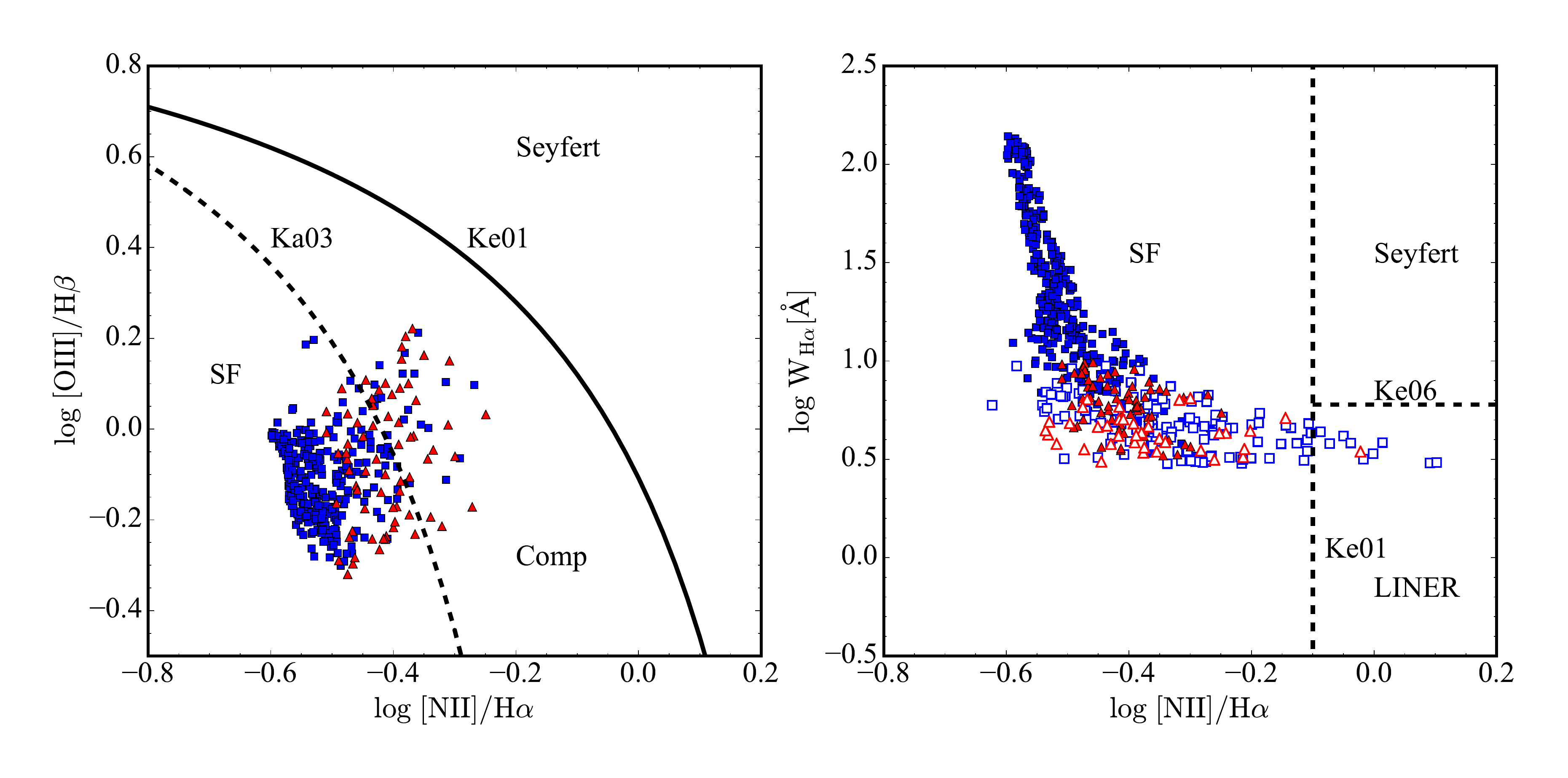}
\caption{Left panel: The traditional BPT diagnostic diagram for SF and TPSB regions. The blue squares and red triangles represent the spaxels with SNR of four emission lines larger than 3 for SF and TPSB, respectively. The dashed and solid lines are from \cite{2003MNRAS.346.1055K} and \cite{2001ApJ...556..121K}, respectively. Right panel: An alternative diagnostic diagram of excitation mechanism for SF and TPSB regions. The hollow squares and triangles represent the spaxels with SNR of H$\beta$ or \oiii\ lower than 3 for SF and TPSB, respectively. The blue solid squares and red triangles are the same as the left panel. The horizontal dashed line is from \cite{2006MNRAS.372..961K}.}
\label{f4}
\end{figure*}

Figure \ref{f3} shows the stacked spectra for SF, TPSB and QPSB regions, respectively. For the spectrum of SF, it shows the strongest H$\alpha$ emission and weakest H$\delta$ absorption. The spectrum of TPSB has similar H$\delta$ absorption but stronger H$\alpha$ emission than that of QPSB. This feature might indicate that PGC 26218 need a period of transition to fade the emission lines from TPSB to QPSB.

\subsection{SFR, Gas-phase Metallicity and $\Sigma_{\rm HI}$}
\label{sec2.4}

The emission lines may be contaminated by active galactic nucleis (AGNs) and shocks, which would lead to the overestimation of star formation rate (SFR) when we use H$\alpha$ as tracer of star formation. To this end, we investigate the excitation mechanisms for SF and TPSB regions according to the traditional BPT diagram (\cite{1981PASP...93....5B}, left panel of Figure \ref{f4}). In this diagram, we impose the SNR of these four emission lines (i.e., H$\alpha$, \nii, H$\beta$ and \oiii) to be greater than 3. For the spaxels with H$\alpha$ and \nii\ SNR larger than 3, we use the relation between the EW (H$\alpha$) and \nii/H$\alpha$ (right panel of Figure \ref{f4}) as the diagnostic diagram of excitation mechanisms \citep{2010MNRAS.403.1036C}. We find no spaxels in the Seyfert region, which indicates that PGC 26218 does not host an AGN. Although some spaxels locate at the composite and LINER regions, the effect from shock is ignorable considering that most of the spaxels locate at the SF region and the contribution from shocks to emission lines could be low in the SF region.
Taking into account that all of the SF and TPSB spaxels have $\rm EW(H\alpha) < -3\ \AA$ and the percent light contribution (at $\lambda = 5100 \AA$) of young populations in those spaxels is larger than 4\%, we believe that most of the contribution is dominated by ionization due to young star and only a tiny fraction by HOLMES (hot low-mass evolved stars)/postAGB stars \citep{2016A&A...590A..44G, 2018MNRAS.474.3727L}. At any rate, our SFR value is an upper limit.

\begin{figure*}
\centering
\includegraphics[angle=0,width=0.95\textwidth]{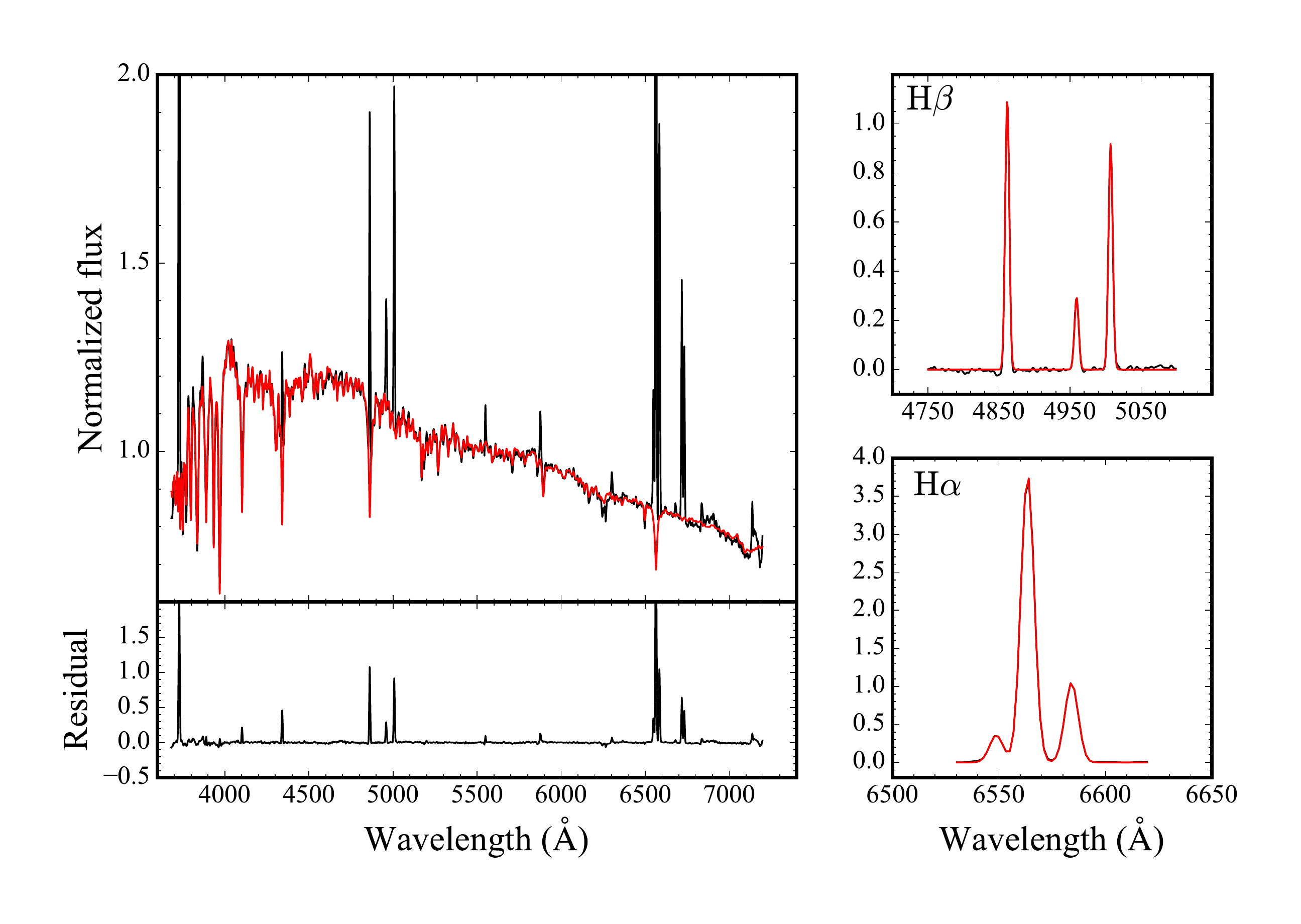}
\caption{Left panels: Results of pPXF 150 simple stellar population synthesis for the stacked spectrum of PGC 26218. The black and red curves are the original and the model spectra, respectively. The residual spectrum is shown at the bottom. Right panels: we highlight the single gaussian fitting for H$\beta$ and H$\alpha$. The black and red lines represent the original data and the model.}
\label{f5}
\end{figure*}

In order to estimate the spatially resolved SFR from extinction-corrected H$\alpha$ emission line, we calculate the extinction according the Balmer decrement assuming \cite{1989ApJ...345..245C} extinction curve with case B condition for each Voronoi bin. The extinction-corrected SFR is calculated using the formula given by \cite{1998ARA&A..36..189K}:
\begin{equation}
SFR(\msun yr^{-1})=7.9 \times 10^{-42} L(\rm H\alpha),
\label{e1}
\end{equation} 
where $L$(H$\alpha$) is the extinction-corrected H$\alpha$ luminosity. For the total extinction-corrected SFR, we construct a spectrum by stacking all spaxels in Voronoi bins of SF and TPSB regions. We use the pPXF code again to model the stellar continuum. For the continuum-subtracted emission-line spectrum, we simultaneously fit H$\alpha$, \mbox{\nii,} H$\beta$, and \oiii\ lines using the single Gaussian model (see Figure \ref{f5}) and calculate the extinction. It is worth noting that the MPA/JHU SDSS DR7 catalog \citep{2004MNRAS.351.1151B} provided the aperture-corrected SFR of PGC 26218 (SFR $\sim$ 0.20 $M_{\odot} \rm yr^{-1}$) based on \cite{2001MNRAS.322..231K} IMF, whereas we adopt \cite{1955ApJ...121..161S} IMF and the total SFR is $\sim$ 0.28 $M_{\odot} \rm yr^{-1}$. The conversion between the two IMF is Kroupa IMF $\times$ 1.5 $\approx$ Salpeter IMF. Therefore our SFR is consistent with that given by \cite{2004MNRAS.351.1151B}.

The inclination and axial ratio (see Table \ref{tab2}) derived from $GALFIT$ \citep[Version 3.0.5,][]{2002AJ....124..266P} show that the projection effect may affect the accuracy of surface densities. We calculate the inclination based on  
\begin{eqnarray}
i=arccos\sqrt{\frac{q^2-q^{2}_{0}}{1-q^{2}_{0}}}
\label{e2}
\end{eqnarray}
where q = b/a, the ratio of minor semi-axis and major semi-axis. We adopt $q_{0} = 0.25$ \citep{1970ApJ...160..831S} instead of 0.2 \citep{1926ApJ....64..321H} to correct the inclination for classical S0 galaxies. We find the inclination-corrected star formation rate surface density ($\Sigma_{\rm SFR}$), also referred to as the intensity of the star formation, is lower by 0.3 dex than that of non-inclination-corrected. However, the correction for the inclination will lead to the equal decrease in molecular gas surface density ($\Sigma_{\rm gas}$) . Thus, the inclination correction will only cause the parameters to move to the lower left in K-S law, but will not significantly affect the results shown in Figure \ref{f9}.

The gas-phase metallicity of PGC 26218 is derived based on the 
Bayesian method\footnote{http://users.obs.carnegiescience.edu/gblancm/izi} \citep{2015ApJ...798...99B} with the \cite{2010AJ....139..712L} photoionization model, which allows us to input arbitrary sets of strong nebular emission lines to infer the probability density functions of gas-phase metallicity and ionization parameter. We use the spectra that is used to compute SFR to estimate the gas-phase metallicity, which is $\sim$ 1 $\sigma$ below the relation derived by \cite{2004ApJ...613..898T} while consistent with the relation given by \cite{2013A&A...554A..58S} within the error range.

We use the average gas-phase metallicity (see Table \ref{tab2}) to derive the surface mass density of atomic gas ($\Sigma_{\rm HI}$) according to the empirical formula given by \cite{2018ApJ...862..110S}. In this scaling relation, the optically thin conversion is adopted and a factor of 1.36 is taken into account to include the heavy elements. Furthermore, the effects of the diffuse \hi\ on the saturation column density is neglected. However, the assumption does not significantly affect the $\Sigma_{\rm HI}$.

\section{NOEMA millimeter Observation and Data Analysis}
\label{sec3}

\subsection{NOEMA millimeter Observation}
\label{sec3.1}

PGC 26218 was observed with NOEMA located in the South of the French Alps on 2018 June 16 (Project S18BN001. PI: Xue Ge). The source was observed for 2.5 hours with the compact D configuration. This configuration is best suited for deep integration and coarse mapping (resolution $\sim 3.7\arcsec$ at 100 GHz and $\sim 1.6\arcsec$ at 230 GHz) and provides the lowest phase noise and highest sensitivity. The source 3C273 is chosen to calibrate the bandpass, while the sources J0851+202 and J0923+282 are chosen to calibrate the phase.
CO($J$=1-0) rest frequency is 115.271 GHz (the redshifted frequency is 114.643 GHz). We only observed the emission line with the receivers in 3 mm band tunable sky frequency between 70.4 and 119.9 GHz. The receiver band has dual-polarization capabilities and each of the two polarizations delivers a bandwidth of 7.744 GHz in the lower sideband (LSB) and upper sideband (USB) simultaneously. Each sideband contains two adjacent basebands of $\sim$ 3.9 GHz width, called inner and outer baseband. The spectral resolution is $\sim$ 2 MHz throughout the wide sidebands.

\begin{figure*}
\centering
\includegraphics[angle=0,width=0.45\textwidth]{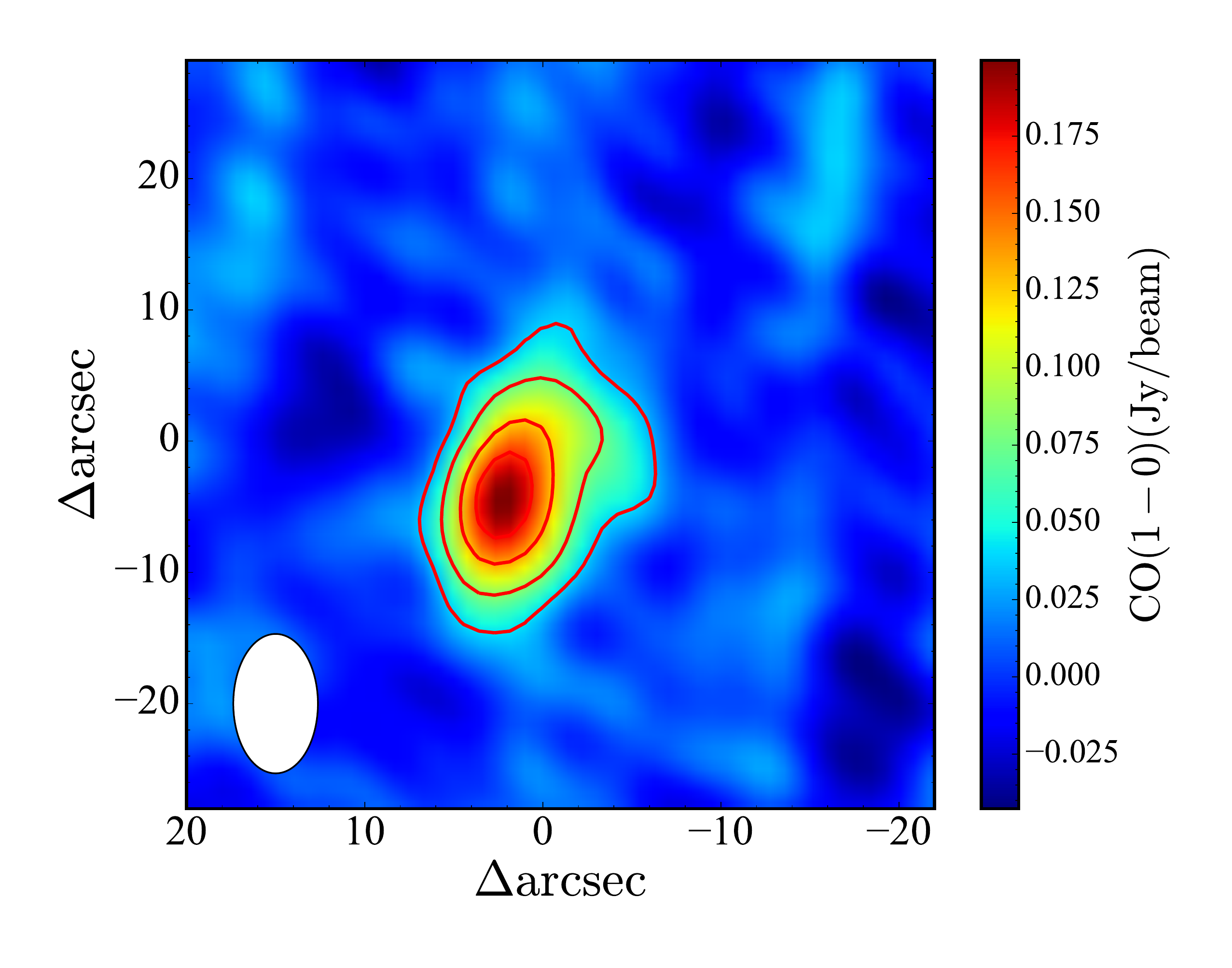}
\includegraphics[angle=0,width=0.45\textwidth]{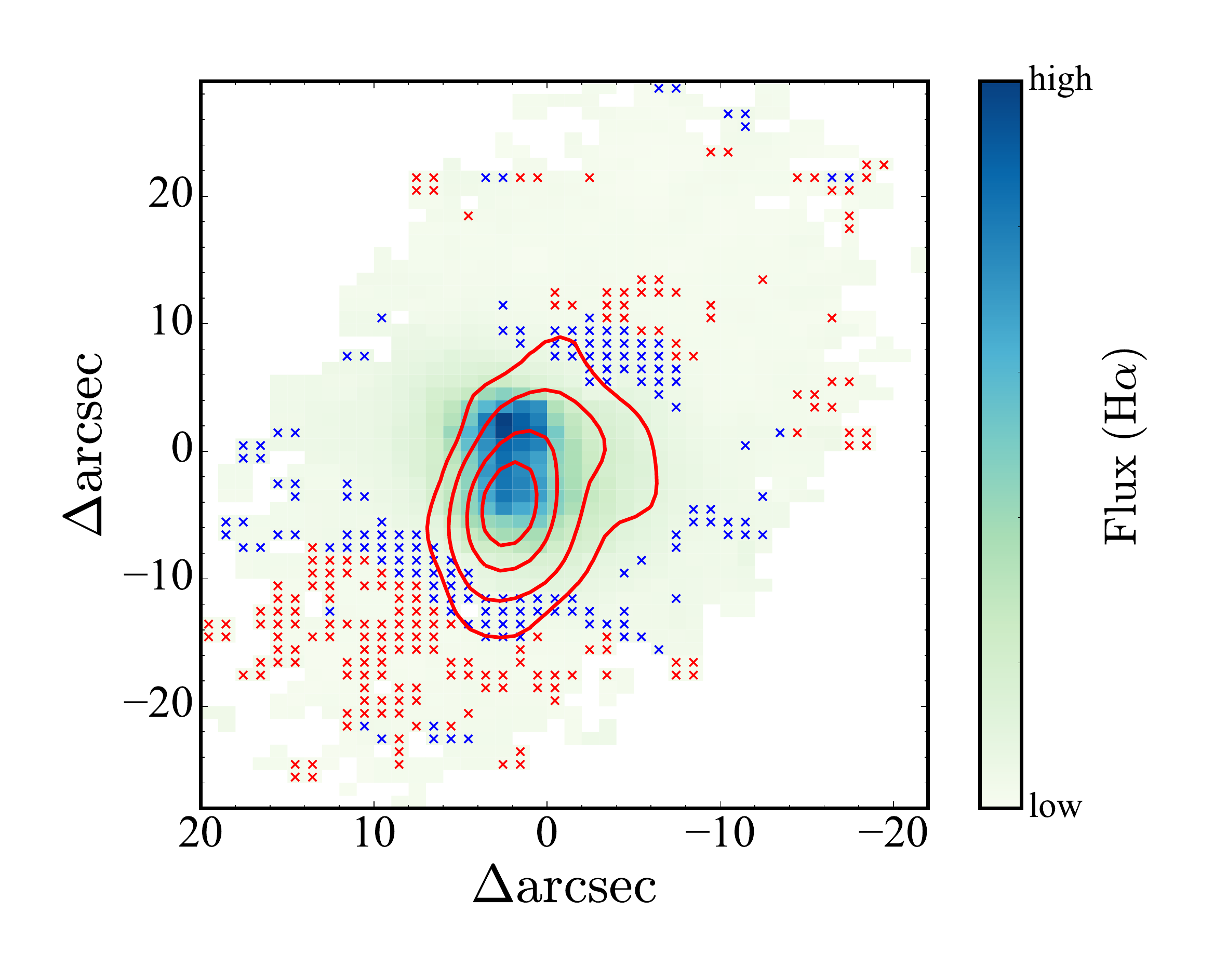}
\caption{Left panel: the image of NOEMA CO($J$=1-0). The synthesized beam size is 10.62 $\times$ 4.76 arcsec as marked in the lower-left corner. Right panel: image of H$\alpha$ overlaid with NOEMA CO($J$=1-0) contours. The levels of the contours are 0.04, 0.08, 0.13, 0.17, respectively. The red and blue crosses are the same as Figure \ref{f2}. The reference coordinates of the two panels have been set in the same scale to show clearly the relative positions of H$\alpha$ and molecular gas regions.}
\label{f6}
\end{figure*}

\subsection{The Data Reduction and $M_{\rm H_{2}}$}
\label{sec3.2}

The CO($J$=1-0) line data was calibrated by $CLIC$, a module of the available $GILDAS$ software package. We choose the channels from -860 $\rm km\ s^{-1}$ to 400 $\rm km\ s^{-1}$, which is wide enough to analysis the CO($J$=1-0). The final channels are smoothed to 10 $\rm km\ s^{-1}$. The cleaning of line image was done by the mapping module of $GILDAS$ and the resulting channel map is shown in the left panel of Figure \ref{f6}. The synthesized beam size is 10.62 $\times$ 4.76 arcsec. The uncertainty of calibration in the flux is $\sim$ 20\% and the 1 $\sigma$ uncertainty is about $\sim$ 5.6 mJy/beam. The reference center of the map is set to be the same as optical center so that we can compare the relative positions of H$\alpha$ and CO($J$=1-0) (see the right panel of Figure \ref{f6}). We can see that the emission of CO($J$=1-0) is mainly concentrated at SF and TPSB regions and basically overlapped with the emission of H$\alpha$. We do not detect CO($J$=1-0) emission in the QPSB regions, but weak emission exists in the TPSB region. In addition, we find that there is an offset between the peaks of optical and the sub-millimeter CO emission (see detailed description in Sec \ref{sec4.4}). 
 
We extract the CO($J$=1-0) emission of each spaxel with enough high SNR (3 $\sigma$) at the peak and construct a stacked spectrum (Figure \ref{f7}). The interesting point is that the CO($J$=1-0) line shows the asymmetry in the profile. We fit the spectrum with two Gaussians. One of them represents the redshift component, while the other represents the rotation component.
It is found that one of the Gaussians shows a redshift of $\sim$ 46 $\rm km\ s^{-1}$, which might indicate the inflow of gas, while the other does not. The asymmetrical profile does not disappear even though we stack the spectra within an effective radius.
The flux density of CO($J$=1-0) is estimated from the best-fit model. We use the relation given by \cite{2005ARA&A..43..677S} to calculate the luminosity of CO($J$=1-0).
\begin{equation}
L^{\prime}_{\rm CO} = 3.25 \times 10^7 \,S_{\rm CO} \Delta v \,\nu_{\rm obs}^{-2}\,D_L^2\, (1+z)^{-3}
\end{equation}
where $S_{\rm CO} \Delta v$, $\nu_{\rm obs}$, and $D_L$ are the CO integrated flux density in units of Jy $\rm km\ s^{-1}$, the observing frequency in GHz, and the luminosity distance in Mpc, respectively. 

The mass of molecular hydrogen ($M_{\rm H_{2}}$) is estimated using following formula:
\begin{equation}
M_{\rm H_{2}}= \alpha_{\rm CO} \times L^{\prime}_{\rm CO}
\end{equation}
where $\alpha_{\rm CO}$ is the conversion factor between CO($J$=1-0) and $\rm H_{2}$. Combining the $\Sigma_{\rm HI}$ derived in Section \ref{sec2.4}, the inclination corrected $\Sigma_{\rm gas}$ ($\Sigma_{\rm HI + H_{2}}$) is estimated by adopting $\alpha_{\rm CO} = 4.3 M_{\sun} (\rm K\ km\ s^{-1} pc^{2})^{-1}$, for the inner disk of Galaxy \citep{2013ARA&A..51..207B}. 
\begin{center}
\begin{table}[h]
\begin{footnotesize}
\renewcommand{\thetable}{\arabic{table}}
\caption{A summary of the physical properties for PGC 26218}
\label{tab2}
\begin{tabular}{lr}
\tablewidth{0pt}
\hline
\hline
PGC 26218		&								\\
\hline
$\rm A_{v}$ [mag]						&	$0.76 \pm 0.06$ \\
$\log$ SFR [$M_{\sun}$ yr$^{-1}$]						&	$-0.55 \pm 0.01$\\
$\log$ $\Sigma_{\rm SFR}$ [$M_\odot\ \rm yr^{-1}\ kpc^{-2}$]                        &        $-1.39 \pm  0.01$ \\
inclination [deg]						&	64.5 \\
$\rm b/a$					&	0.49 \\
$12+\rm \log\ (O/H)$ 						&	$8.53 \pm  0.13$ \\
$S_{\rm CO (J=1-0)}$ $\Delta v$ [Jy km s$^{-1}$]		&	6.78 $\pm$ 0.75 \\
$\log\,L^{\prime}_{\rm CO (J=1-0)}$ [K km s$^{-1}$ pc$^2$]	&	$6.96 \pm  0.75$   \\
$\log$ $\rm M_{H_{2}}$ [$M_{\sun}$]						&	$7.60 \pm  0.15$ \\
$\log$ $\Sigma_{\rm gas}$ [$M_\odot\ \rm pc^{-2}$]                        &        $1.30 \pm  0.14$ \\
$\log$ $\Sigma_{1}$ [$M_{\sun}\ \rm kpc^{-2}$]						&	8.25 \\
\hline
\end{tabular}
\end{footnotesize}
\end{table}
\end{center}


\section{Results and Discussions}
\label{sec4}

\subsection{The Star Formation and Metallicity in PGC 26218}
\label{sec4.1}

\begin{figure}
\centering
\includegraphics[angle=0,width=0.5\textwidth]{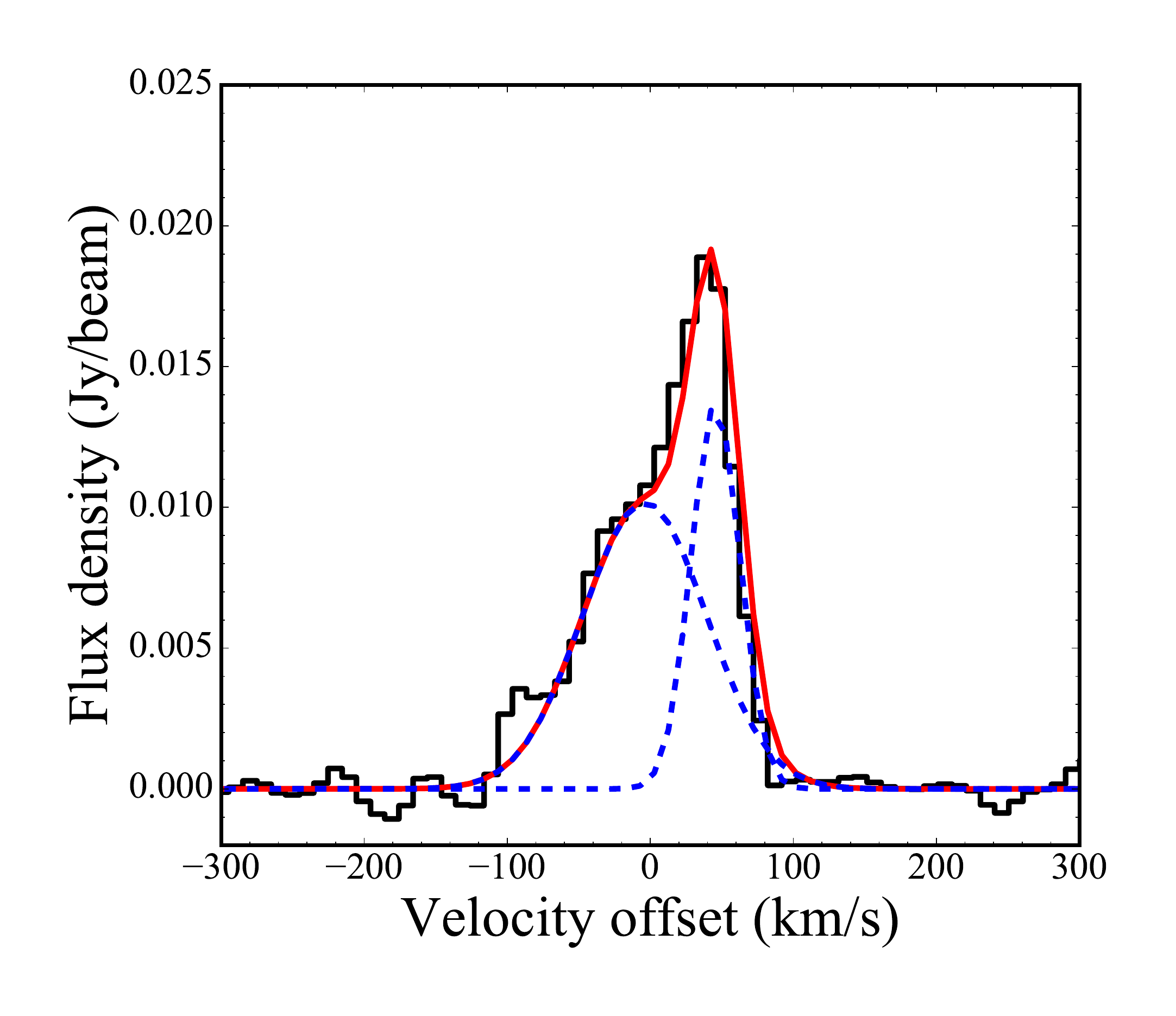}
\caption{The decomposition of CO($J$=1-0) profile, smoothed to a resolution of 10 $\rm km\ s^{-1}$. The black line represents the original spectrum and the blue dashed lines Gaussian models. The red solid line is the best-fit model, the superposition of the two Gaussian profiles.} 
\label{f7}
\end{figure}

\begin{figure}
\centering
\includegraphics[angle=0,width=0.55\textwidth]{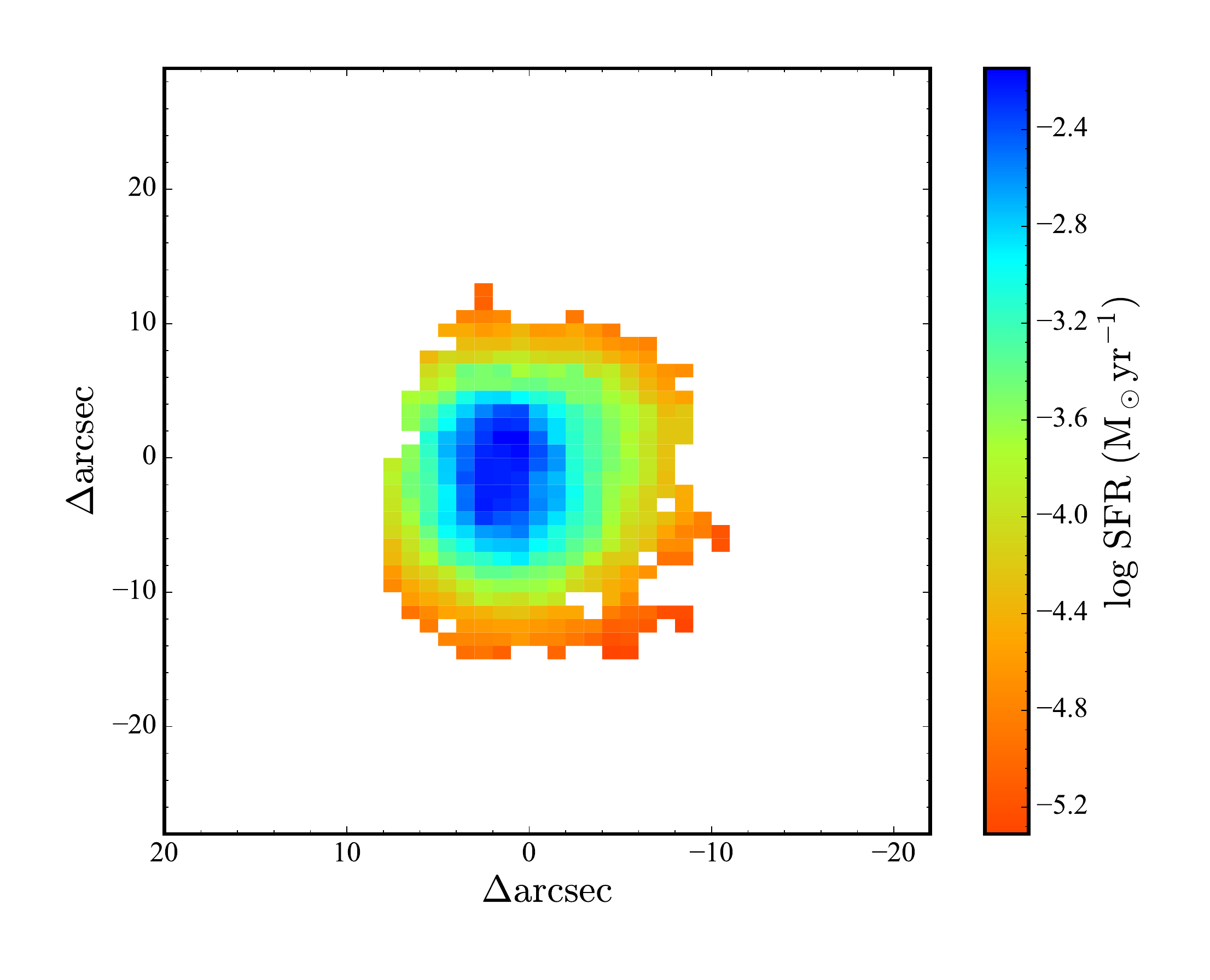}
\caption{The spatial distribution of SFR for PGC 26218.}
\label{f8}
\end{figure}

Figure \ref{f8} shows the map of SFR in logarithmic space for PGC 26218. It is found that star formation occurs mainly in an effective radius, which is consistent with the distribution of CO emission. As mentioned in Figure \ref{f7}, the component of redshift might indicate the gas inflow, which could have led to the star formation in the central region.

Previous studies have shown that the stellar mass surface density within the central 1 kpc, $\Sigma_{1}$, strongly correlates with the star formation \citep{2012ApJ...760..131C, 2013ApJ...776...63F}. In order to compare our result to \cite{2013ApJ...776...63F}, we also adopt the stellar mass from MPA-JHU DR7 catalog (e.g., $\log M_{\ast}=9.15$) and similar method to estimate the $\Sigma_{1}$. Here, we estimate the $\Sigma_{1}$ from the NASA-Sloan Atlas catalog providing the surface brightness profiles in series of angular sizes. The extinction-corrected and k-corrected surface brightness profiles within 1 kpc in $i$-band are used to convert into the $\Sigma_{1}$ according to the mass-to-light ratio of $i$-band (see \cite{2013ApJ...776...63F} for detailed calculation). The value of $\log$ $\Sigma_{1}$ for PGC 26218 is 8.25 $M_\odot\ \rm kpc^{-2}$ (see Table \ref{tab2}), which is 0.34 dex ($\sim$ 2 $\sigma$) below the best fitting relation of $\Sigma_{1}$-$M_{\ast}$ obtained from the combination of blue and green valley galaxies \citep{2013ApJ...776...63F}.
We also extract the mass value within 1 kpc from the stellar population analysis method and find that the value of $\Sigma_{1}$ does not change significantly. The position of PGC 26218 in $\Sigma_{1}$-$M_{\ast}$ relation suggests that the central star formation might not be suppressed significantly. With the current SFR, the remaining molecular gas can last for about 0.1 Gyr.

\subsection{SFMS and K-S Law}
\label{sec4.2}

The star forming main sequence (SFMS) \citep{2007A&A...468...33E, 2007ApJS..173..267S} shows the relationship between stellar mass and SFR. S0 galaxies are often thought to be the absence of noticeable star formation regions. To investigate whether PGC 26218 resides in the SFMS, we put the source on the SFMS relation given by \cite{2007A&A...468...33E}, who obtained SFR and stellar mass from the MPA-JHU DR4 catalog\footnote{http://www.mpagarching.mpg.de/SDSS/DR4/.}. It is noted that the SFMS in \cite{2007A&A...468...33E} was calibrated based-on \cite{1955ApJ...121..161S} IMF. So, we adopt here $\log M_{\ast}=9.32$ to investigate the position of PGC 26218 in the SFMS. We find PGC 26218 deviates slightly from the best fitting relationship (slightly below SFMS), but it follows the SFMS relation well within the error range. \cite{2014ApJS..214...15S} gave a redshift-dependent SFMS relation using a compilation of many studies from the literature. We find that PGC 26218 still follows the redshift-dependent SFMS relation considering 0.2 dex scatter.

\begin{figure}
\centering
\includegraphics[angle=0,width=0.45\textwidth]{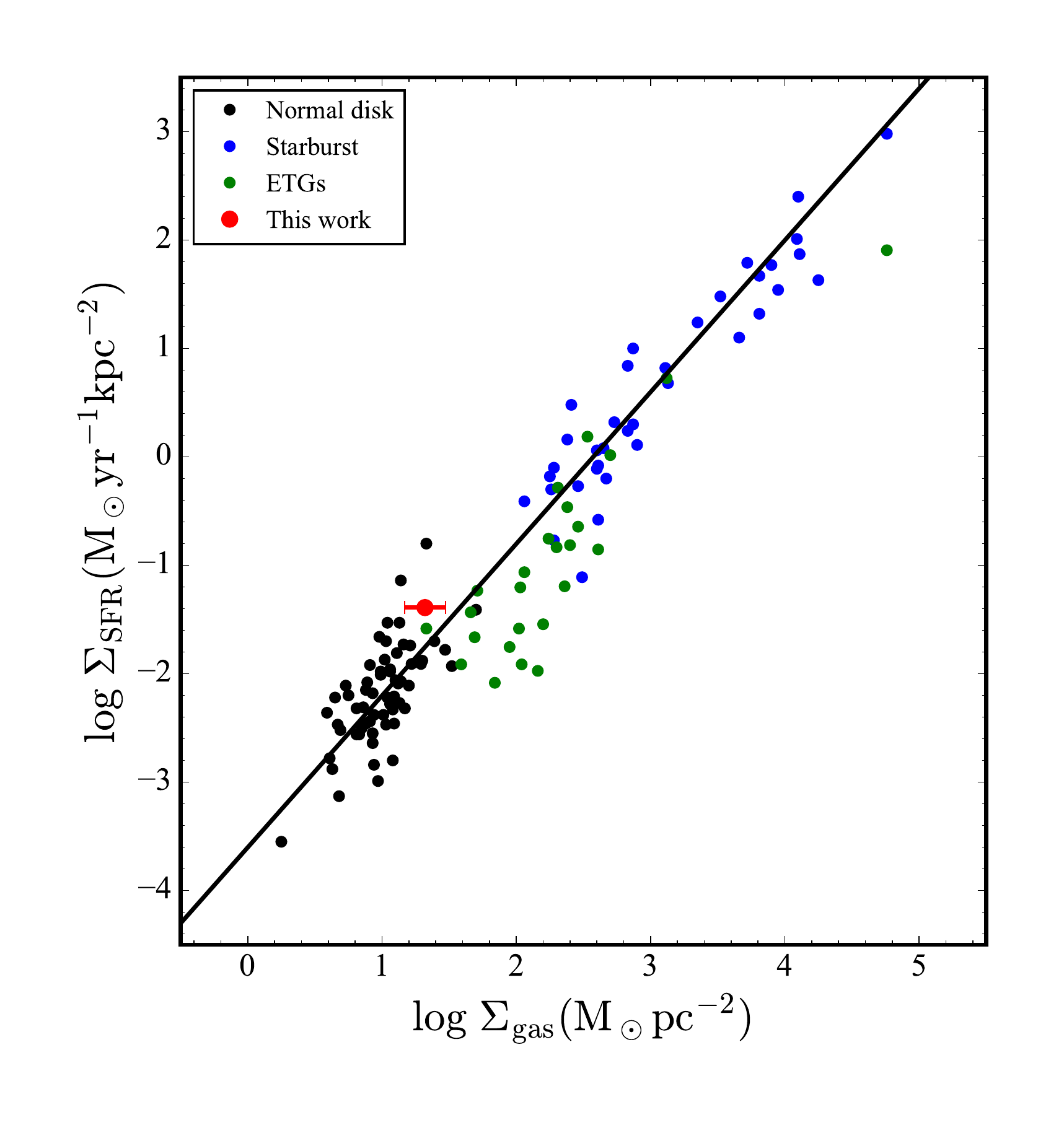}
\caption{The relationship between $\log$ $\Sigma_{\rm gas}$ and $\log$ $\Sigma_{\rm SFR}$. The green points represent the $\rm ATLAS^{3D}$ ETGs with spatially resolved CO($J$=1-0) detection from \cite{2014MNRAS.444.3427D}, while black and blue points represent the normal and starburst galaxies form \cite{1998ApJ...498..541K}. The solid line is the best fitting for \cite{1998ApJ...498..541K} sample. PGC 26218 is marked in red.}
\label{f9}
\end{figure}

The tight relationship between $\Sigma_{\rm gas}$ and $\Sigma_{\rm SFR}$ (i.e., K-S law) has been found in different types of star-forming galaxies \citep{1959ApJ...129..243S, 1998ApJ...498..541K, 2012ARA&A..50..531K}. Figure \ref{f9} displays the K-S law for ETGs, star-forming and starburst galaxies. We find that the ETGs (green points, most of them are S0 galaxies) in \cite{2014MNRAS.444.3427D} deviates from this relationship, while PGC 26218 (red point) basically obeys the K-S law, which implies that some mechanisms enhance its SFR. It is noted that the $\Sigma_{\rm SFR}$ is computed in the same region as the CO($J$=1-0) measurement (see Sec \ref{sec3.2}). We find that the result will not change if we compute the $\Sigma_{\rm SFR}$ and $\Sigma_{\rm gas}$ in an effective radius. Although the fraction of molecular gas for PGC 26218 ($\sim$ 2\%) is slightly lower than that of \cite{2014MNRAS.444.3427D} ($\sim$ 3\%), the star formation efficiency in PGC 26218 might be higher.
\cite{2010ApJ...725L..62W} presented the relationship between molecular gas and star formation in low-mass ellipticals/S0 galaxies. They found that most blue-sequence ellipticals/S0 galaxies show higher $\Sigma_{\rm SFR}$ at fixed $\rm M_{H_{2}}$, which is similar to local dwarf galaxies. In addition, the fraction of molecular gas is a factor of 2 higher than that of PGC 26218. \cite{2017A&A...605A..74K} studied the star formation of the same sample as \cite{2014MNRAS.444.3427D} in 
$\rm ATLAS^{3D}$ survey. They found that the local ETGs have the similar star formation efficiency to the star-forming galaxies and follow the K-S law. The difference between them might be attributed to the methods of calculating SFR. \cite{2014MNRAS.444.3427D} calculated the SFR via FUV and 22 $\mu$m, while \cite{2017A&A...605A..74K} via the spectral energy distributions fitting. 
\cite{2018MNRAS.475.1791C} combined the CO observations made by the Combined Array for Millimeter-wave Astronomy from Extragalactic Database for Galaxy Evolution survey \citep{2017ApJ...846..159B} with the 2D-spectroscopic observations made by CAHA from Calar Alto Legacy Integral Field Area survey \citep{2012A&A...538A...8S}. They calculated the $\Sigma_{\rm SFR}$ from the 
H$\alpha$ map and investigated the dependence of $\Sigma_{\rm SFR}$ on the Hubble types. They found that S0 galaxies have lower $\Sigma_{\rm SFR}$ than that of spirals, which is inconsistent with PGC 26218.

\subsection{Kinematics Of Star and Gas}
\label{sec4.3}

\begin{figure*}
\centering
\includegraphics[angle=0,width=0.3\textwidth]{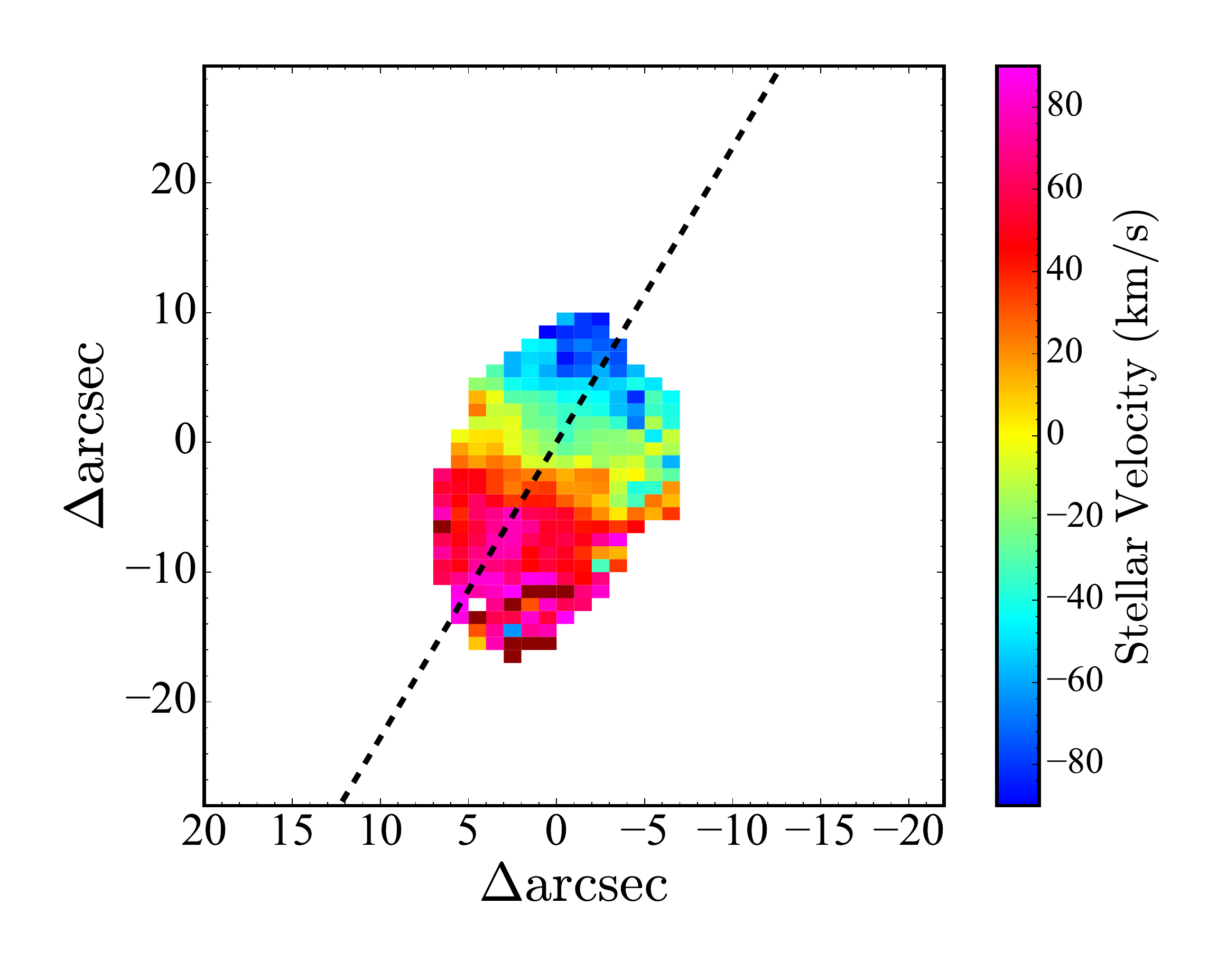}
\includegraphics[angle=0,width=0.3\textwidth]{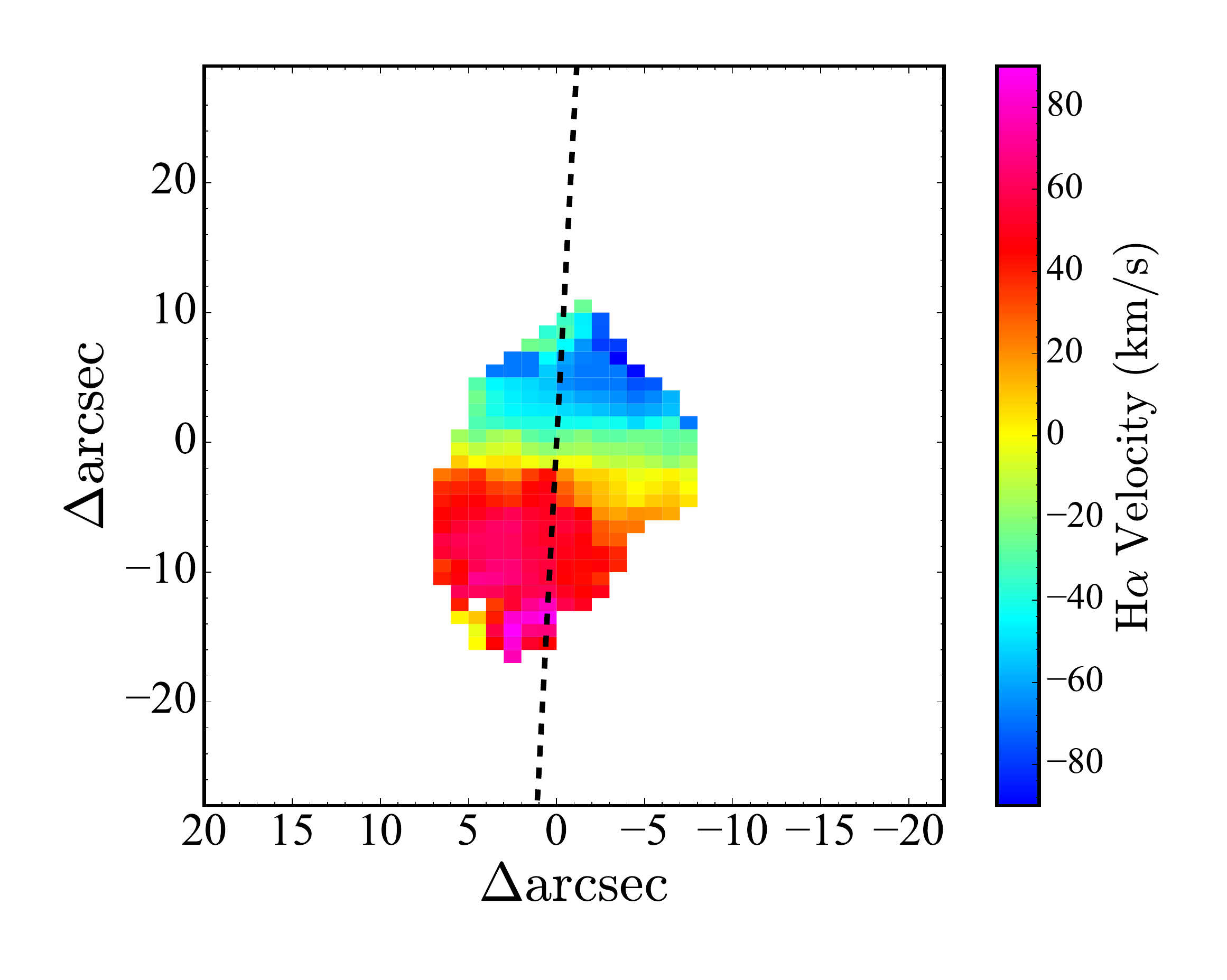}
\includegraphics[angle=0,width=0.3\textwidth]{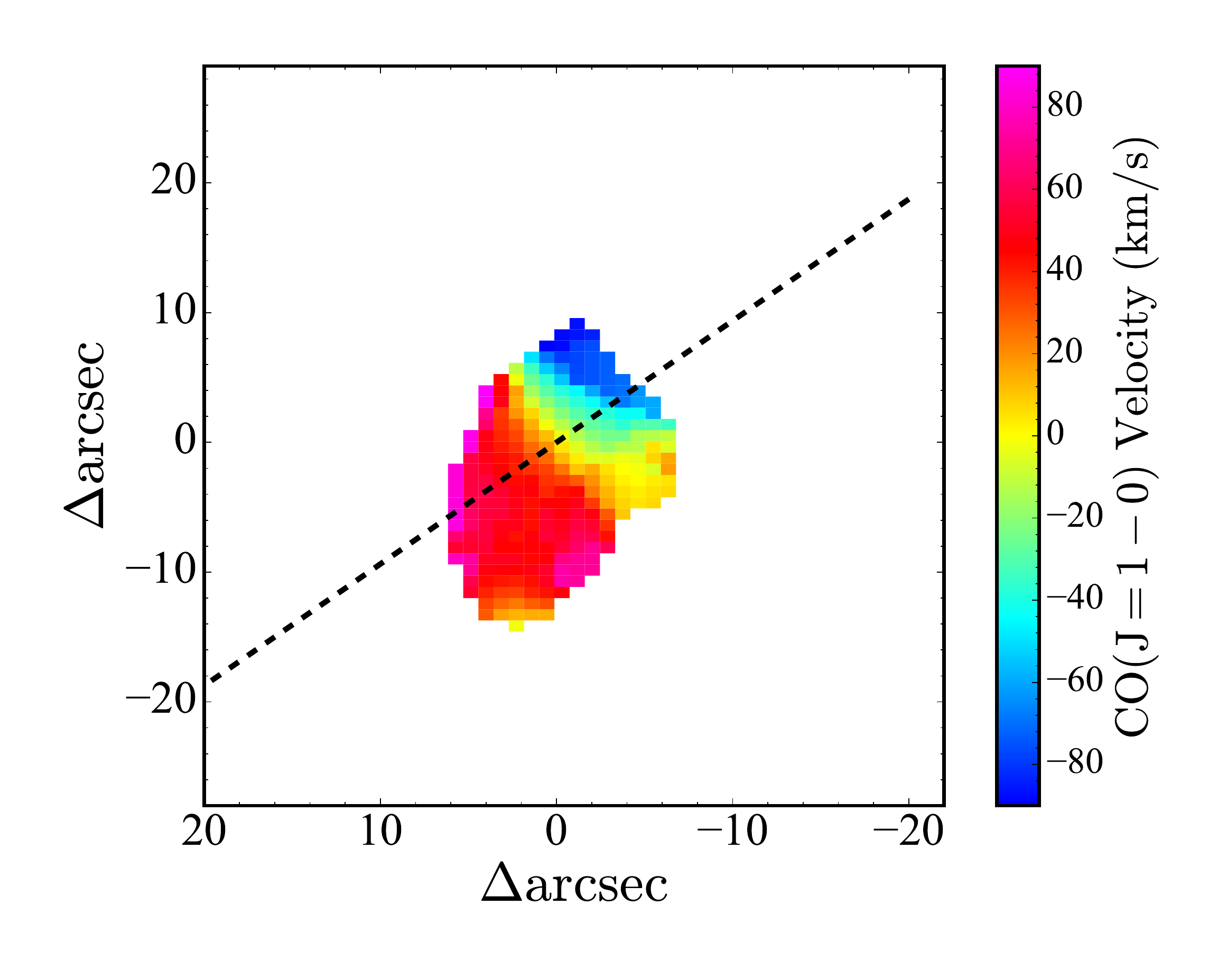}
\caption{The maps of velocities for star (left panel), ionized gas (middle panel), and CO($J$=1-0) (right panel). The dashed lines represent the major-axis of rotation.}
\label{f10}
\end{figure*}

According to the region of CO($J$=1-0), we compare the kinematics of star, H$\alpha$ and CO($J$=1-0). Figure \ref{f10} shows the distributions of the line-of-sight velocities for these three components. The major-axis of stellar, H$\alpha$ and CO($J$=1-0) velocity fields (dashed lines of Figure \ref{f10}) are measured using the KINEMETRY code \citep[]{2006MNRAS.366..787K}. We define the position angles as the counter-clockwise angle between north and a line that bisects the velocity field of gas or stars on the receding side. We find that the major-axis of stellar, H$\alpha$ and CO($J$=1-0) velocities are 156 $\pm\ 2^{\circ}$, 178 $\pm\ 4^{\circ}$ and 133 $\pm\ 10^{\circ}$, respectively. The difference of rotation axes between H$\alpha$ and CO($J$=1-0) is larger than 45$^{\circ}$. In addition, we find that the velocity of CO($J$=1-0) displays the asymmetry, which indicates that the gas has a motion relative to the stellar components. In general, the kinematic misalignment between ionized gas and CO($J$=1-0) implies that the fuels for the star formation might come from external environment.

\subsection{Multiple Nuclear Structures and the Origins of Star Formation}
\label{sec4.4}

\cite{2017A&A...598A..45G} found that star-forming ellipticals/S0 galaxies display the features of the gas-rich minor merger that fuel the central star formation. 
The above results give us the motivation to explore the mechanisms causing PGC 26218 to undergo a burst of star formation.

\begin{figure*}
\centering
\includegraphics[angle=0,width=0.3\textwidth]{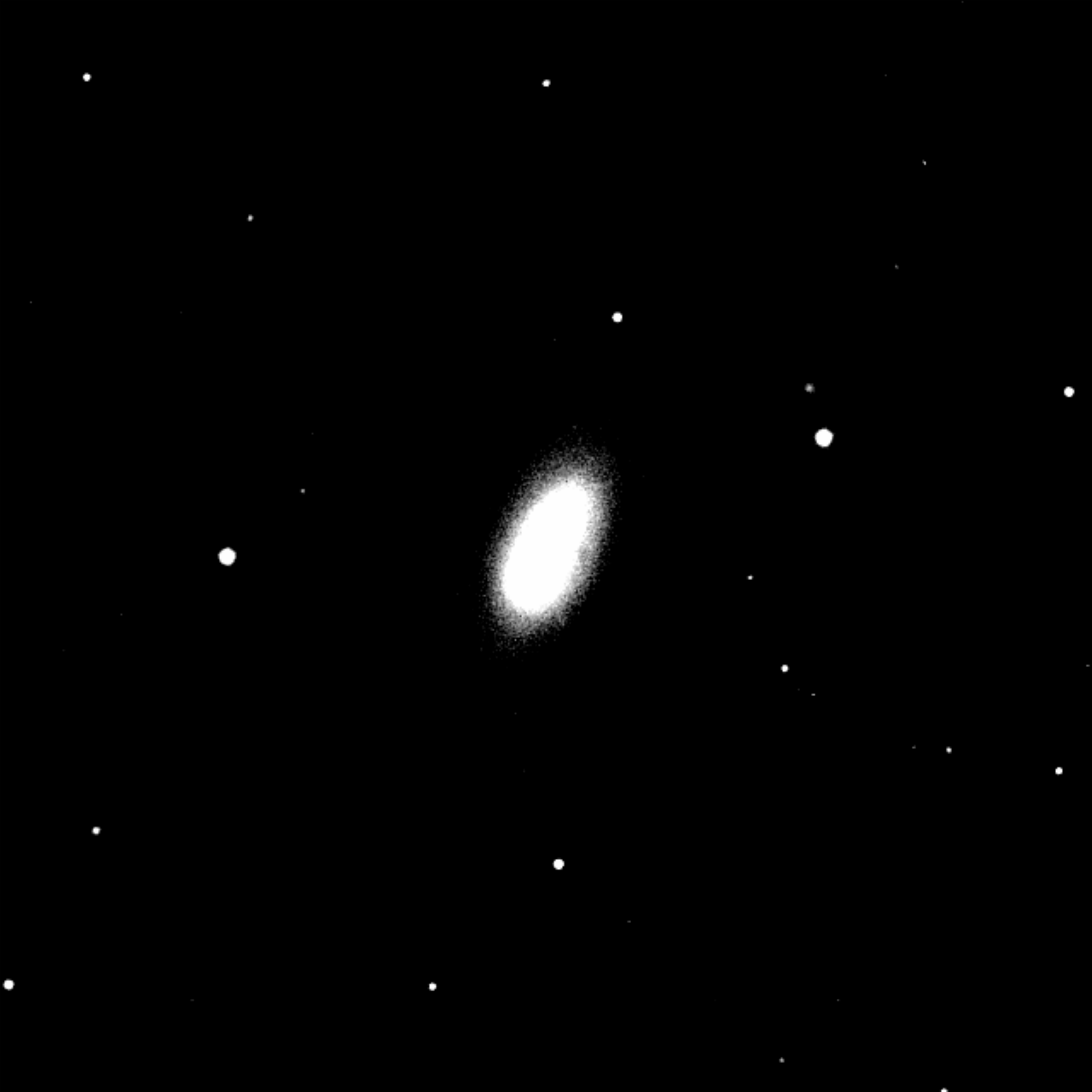}
\includegraphics[angle=0,width=0.3\textwidth]{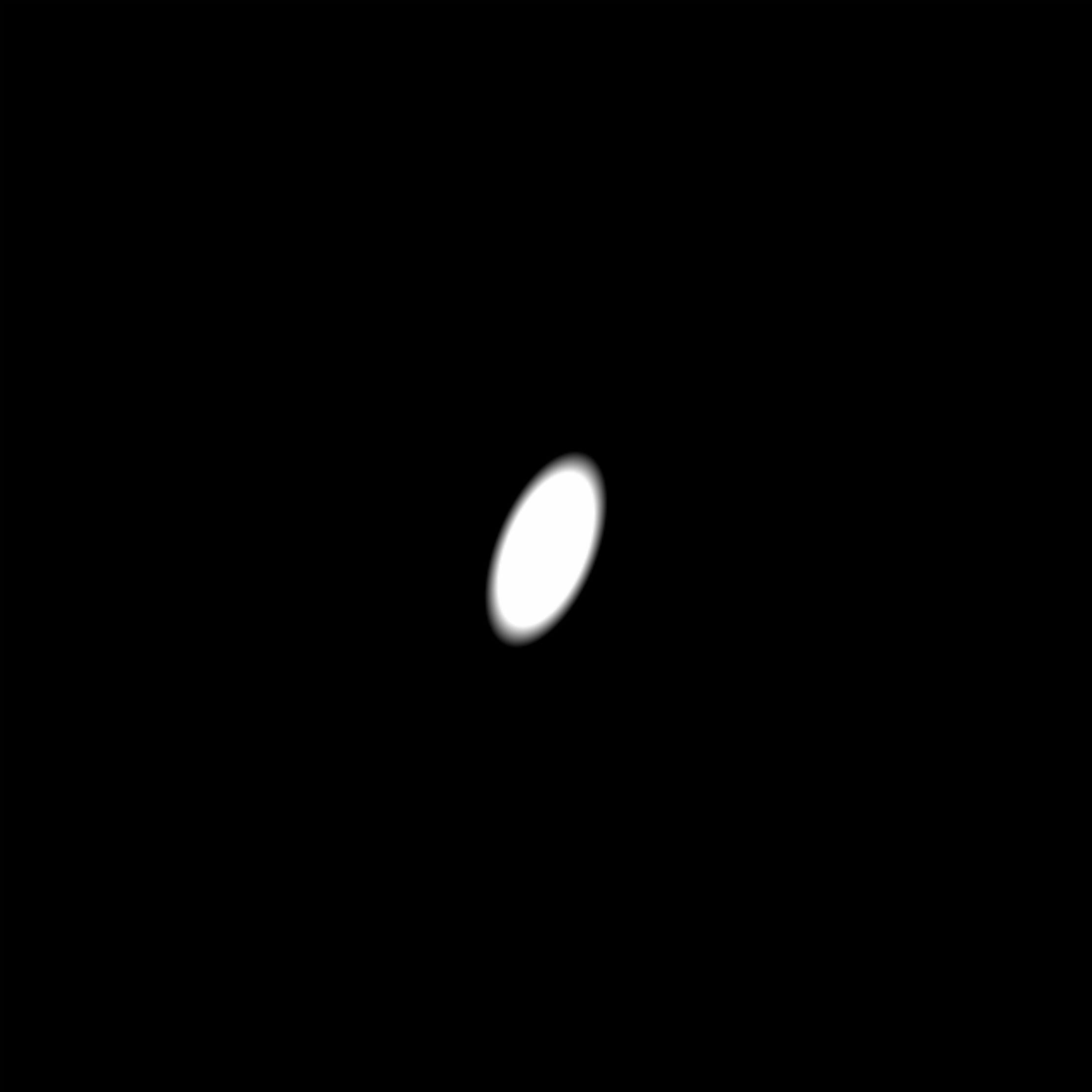}
\includegraphics[angle=0,width=0.3\textwidth]{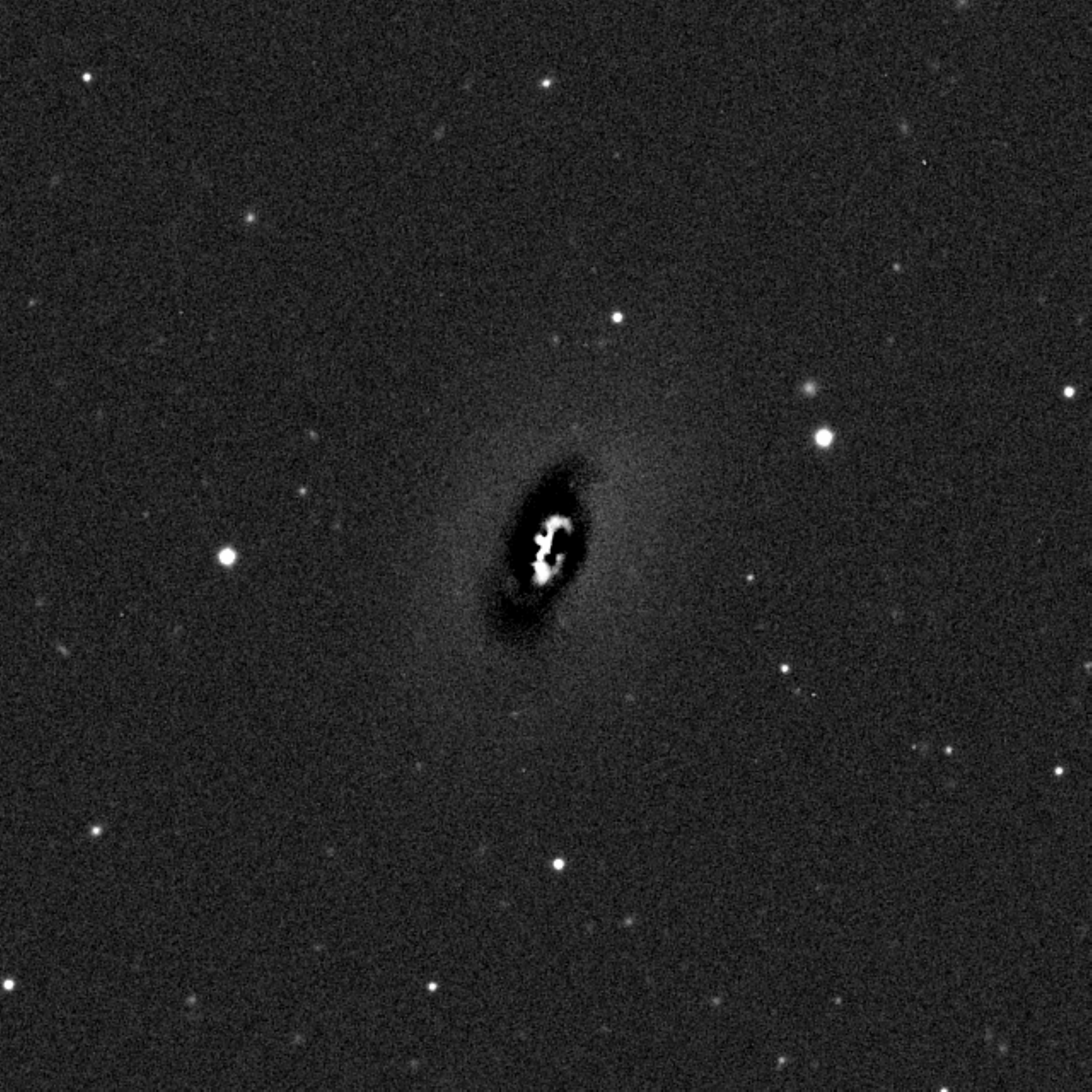}
\caption{The $GALFIT$ fitting for only disk component in SDSS $r$-band. Data image, model image and residual image are shown from left panel to right panel, respectively.}
\label{f11}
\end{figure*}

\begin{figure}
\centering
\includegraphics[angle=0,width=0.5\textwidth]{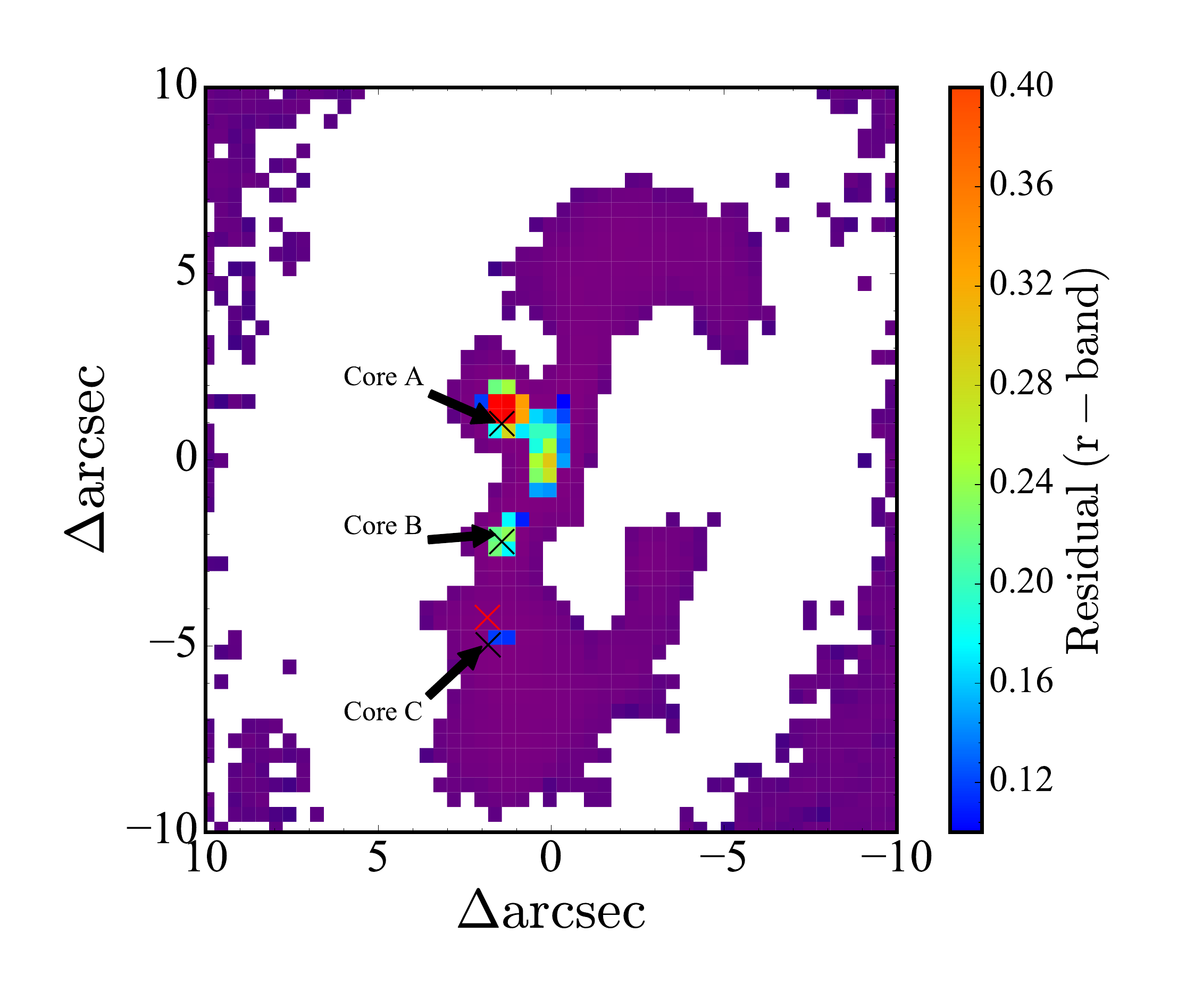}
\caption{The same residual image as Figure \ref{f11}. The center of the image is the same as Figure \ref{f2}. Three black crosses represent the positions of cores defined by us and the red cross indicates the center of CO($J$=1-0). The region is marked in purple if the residual is lower than 0.1.}
\label{f12}
\end{figure}

The morphological disturbation of galaxies can give us clues into the origins of nuclear activities. In order to investigate the central structure of PGC 26218, we simply use a disk component (s\'ersic index = 1) to model the galaxy disk of PGC 26218 in $r$-band (Figure \ref{f11}). We can clearly see from the residual image that PGC 26218 displays a peculiar structure at its center. It is worth noting that the peculiar structure in the residual image does not disappear even though we use two components (i.e., bulge and disk components) to decompose the $r$-band image. The s\'ersic indexes of bulge and disk are 1.2 and 1.0, respectively. In order to highlight the structure of the central region, we compute only the disk component assuming a pure disk model. Figure \ref{f12} highlights the disk-subtracted residual image. We find PGC 26218 shows the extreme morphological peculiarities that may be triggered by the galactic merger. We mark the positions with the highest flux compared surroundings as core a, core b and core c in $r$-band, respectively. 
It is found that the projected distances between core a and core b, core a and core c, and core b and core c are $3.2\arcsec$ (0.35 kpc), $6.0\arcsec$ (0.65 kpc), and $2.8\arcsec$ (0.31 kpc), respectively. The peak of CO($J$=1-0) is nearly overlapped with core c and it lies $\sim$ 0.6 kpc from the center of H$\alpha$ (i.e., core a). However, we should keep in mind that the offset is comparable to the resolution of CO($J$=1-0). Higher resolution observation is needed to resolve this question.

Actually, many previous studies have presented the presence of multiple optical nuclei of PGC 26218.
\cite{1993ApJS...85...27M} presented the properties of over 100 Markarian galaxies with multiple nuclear structures or peculiar morphologies. They used the software, IMAGES, to measure the structure of nuclei and found that PGC 26218 (Mrk 1230 in their paper) shows extended nuclear structure along the major axis. \cite{1995ApJS...99..461N} studied the morphologies and kinematics of 16 Markarian galaxies. Their results suggested that PGC 26218 has three nuclei and the farthest and nearest projected distances between the two nuclei are 1.2 kpc and 0.5 kpc, respectively. In addition, PGC 26218 is also included in \cite{2004AJ....128...62G} as a double nucleus case. They found that the separation between the two nuclei is about 0.4 kpc. We note that if the two nuclei depicted by \cite{2004AJ....128...62G} are a and b components, respectively in our Figure \ref{f12}, then the projected separation estimated by us agrees well with theirs.

There are several scenarios that can be used to explain the occurrence of central star-formation activity.
i) The internal secular evolution, such as the turbulence of bar \citep{1990ApJ...363..391P}. The presence of bar causes the gas to inflow to the center, thus inducing the starburst and fueling the mass concentration. In this scenario, the vigorousness of starburst induced by the bar depends on the stellar mass. \cite{2016MNRAS.463.1074C} suggested that the massive ($M_{*} > 2 \times 10^{10} M_{\sun}$) barred galaxies are easier to consume gas by bar-induced starburst than low-mass barred galaxies. We do not expect that the starburst occurred several hundred Myr ago was triggered by the bar because we do not find a prominent bar component in such a low-mass S0 galaxy.
ii) Disk instability. The instability of the gas-rich disk may lead to the formation of star-forming clumps, which can lose angular momentum due to interactions and fall towards the center of the galaxy \citep{2008ApJ...688...67E, 2009ApJ...703..785D}. The observational evidences show that the dissipational processes triggered by disk instabilities are expected to occur in massive galaxies in earlier cosmic times where there is higher gas fraction than that of local universe \citep{2007ApJ...671..303B, 2010ApJ...713..686D}. However, we do not expect a large amount of disturbed gas to be conserved in such a low-mass, low-$z$ S0 galaxy because the current fraction of molecular gas mass is only $\sim$ 2\% of the stellar mass and the galaxy has relatively regular rotations in stellar and gas disks.
iii) Major and minor mergers. Cosmological hydrodynamic simulations show that the frequency of major mergers declines with the cosmic time \citep{2006ApJ...647..763M}. They found that the average merger rate of massive galaxies is 0.054 $\rm Gyr^{-1}$ at z $\sim$ 0.3, while 0.018 $\rm Gyr^{-1}$ for low-mass galaxies. It is believed that major mergers may be too destructive to preserve the inner components and regular disks. 
Minor mergers are expected to significantly increase the SFR \citep{2012ApJ...758...73S, 2014MNRAS.440.2944K} although they are not as violent as major mergers. In low redshift, the incidence of minor mergers is higher than that of major mergers and the minor mergers can build bulges without destroying the disk structure.
As suggested by \cite{2011A&A...533A.104E}, minor mergers can explain the existence of multiple inner components in unbarred galaxies, although this mechanism is more complex than other processes \citep[such as bars and ovals,][]{2004ARA&A..42..603K}.
\cite{2018A&A...617A.113E} have studied whether galaxy mergers can reproduce the features of S0-like remnants based on the GalMer simulated database. They suggested that the mergers can result in relaxed morphologies and inner subcomponents, such as ovals, lenses and compact sources. Furthermore, the relics are more durable in minor mergers than major mergers. For PGC 26218, the minor merger leads to the slow rearrangement of gas from the disk to the center of galaxy and this progress has not yet been completed in a short time (as the multiple nuclear structures we see).

\cite{2017MNRAS.467.4540B} studied stellar kinematics of four S0 galaxies with $M_{*} \gtrsim 10^{10} M_{\sun}$ utilizing the DEIMOS instrument on the Keck telescope. They found that these S0 galaxies generally resemble the spiral progenitors more than the merger remnants. Although they suggested that these S0 galaxies are likely formed via faded disks, the merger events can not be ruled out for the formation of S0 galaxies.
\cite{2018MNRAS.481.5580F} investigated the relative importance of gas stripping and merger using 279 S0 galaxies from MaNGA (Mapping Nearby Galaxies at APO) survey \citep{2015ApJ...798....7B}. They found that the bulges of low-mass S0 galaxies are almost always younger than their disks, which is consistent with PGC 26218. However, they contributed it to the bulge rejuvenation or disc fading.
\cite{2009ApJ...695....1T} studied the star formation using a sample of local ETGs from Spectroscopic Areal Unit for Research on Optical Nebulae (SAURON) survey \citep{2001MNRAS.326...23B}. They found that the cold gas for star formation in S0 galaxies is created via the stellar mass loss.
Similar result has been found by \cite{2010MNRAS.402.2140S}. Furthermore, they also suggested that the minor mergers might be building up the bulges of red sequence S0 galaxies.
\cite{2009MNRAS.394.1713K} and \cite{2010IAUS..262..168K} studied the importance of minor mergers in low-level star formation of ETGs. They found that the minor mergers are the main mechanism for the driver of star formation in low redshift.
\cite{2019MNRAS.488L..80M} studied the star formation of local S0 galaxies with $M_{*} \gtrsim \times 10^{10} M_{\sun}$ from CALIFA survey. Their results show that the star formation in bulges is significantly lower than that of disks. This contradiction with our result might be contributed to the type of bulge because a larger bulge is more likely to quench the star formation in the central region via the morphological, but related with mass, quenching scenario. This can be also understood by differences of peculiar objects (as their NGC 3773 case) from the general trend of low mass objects ($M_{\ast} < 10^{10} M_{\sun}$) as seen in the dispersion in the derived stellar population properties in the low mass range \citep{2017A&A...608A..27G}.

In summary, PGC 26218 is a good case to connect the relevance between the minor merger and the star formation in the past and present, together with the evolutionary trajectories of S0 galaxies. We suggest by combining the results in this work, together with the previous studies that PGC 26218 could have undergone a gas-rich minor merger, which triggered the starburst at the outskirts a few hundred Myr ago and the central star-formation activity considering the multiple nuclear structures.

\section{Summary}
\label{sec5}

We have studied the physical properties of a nearby S0 galaxy with the features of nuclear star-forming activity and post-starburst outside in the disk. Based on the optical IFU spectroscopic observation from CAHA and the millimeter observation from NOEMA, we have established the possible connection between the starburst happened several hundred Myr ago, nuclear star-formation activity, extreme morphological peculiarity, and minor merger. Our main results are summarized as follows:

1. Based on IFU spectroscopic observations, the S0 galaxy PGC 26128 shows properties of post-starburst at the outskirts with star formation in the central region. The on-going star formation prevents the S0 galaxy from deviating the SFMS. 

2. The SFR and the $\rm log$ $M_{\rm H_{2}}$ of PGC 26218 are 0.28 $M_{\odot} \rm yr^{-1}$ and 7.60 $M_{\odot}$, respectively. With this mass of molecular hydrogen, the star formation of PGC 26218 can last about 0.1 Gyr. PGC 26218 obeys K-S law very well as normal disk galaxies and starburst galaxies. The point is that whether the class of galaxy is frequent, which needs to be further explored with large sample of S0 galaxies in local universe.  

3.  We find that the rotation axes of star, ionized gas and CO($J$=1-0) differ by more than 20$^{\circ}$, respectively. The difference between ionized gas and CO($J$=1-0) reaches 45$^{\circ}$, which indicates that the fuel that provides PGC 26218 to form stars might come from the surrounding environment.

4.  PGC 26218 has multiple nuclear structures in optical SDSS image and the CO($J$=1-0) emission line shows the asymmetric profile (46 km/s redshift component). These results indicate that this galaxy may have undergone a gas-rich minor merger, which triggered the starburst a few hundred Myr ago and swept gas into the center, leading to the star formation.

5. There is an offset between the centers of H$\alpha$ and CO($J$=1-0) with a projected offset of 0.6 kpc. However, the offset should be further investigated by higher resolution observations considering that the offset is comparable to the size of semi-major axis of beamsize.


\section*{Acknowledgements}
This work is supported by the National Key Research and Development Program of China (No. 2017YFA0402703) and by the National Natural Science Foundation of China (No. 11733002). In addition, we acknowledge the supports of the staff from CAHA and NOEMA. 
Rub\'en Garc\'ia-Benito acknowledges financial support from the Spanish Ministry of Economy and Competitiveness through grant 205 AYA2016-77846-P. Rub\'en Garc\'ia-Benito acknowledges support from the State Agency for 206 Research of the Spanish MCIU through the ``Center of Excellence Severo Ochoa'' award to 207 the Instituto de Astrof\'isica de Andaluc\'ia (SEV-2017-0709).






\end{document}